\documentclass[12pt]{article}

\usepackage{amsfonts}
\usepackage{amsmath}
\usepackage{amsthm}
\usepackage{amssymb}
\usepackage[doublespacing]{setspace}
\usepackage{graphicx}
\usepackage{rotating}
\usepackage{pdflscape}
\usepackage{authblk}
\usepackage{color}
\usepackage{xcolor,colortbl}
\usepackage[round]{natbib}
\usepackage{hyperref}
\usepackage{caption}
\usepackage{subcaption}
\usepackage[]{algorithm2e}
\usepackage{multirow}
\usepackage{pdflscape}
\usepackage[utf8]{inputenc}
\usepackage{ulem} 
\usepackage{natbib}
\usepackage{hyperref}

\definecolor{light-gray}{gray}{0.8}

\allowdisplaybreaks[1]
\usepackage[flushleft]{threeparttable}

\definecolor{dkgreen}{rgb}{0,0.4,0}
\definecolor{gray}{rgb}{0.5,0.5,0.5}
\definecolor{mauve}{rgb}{0.58,0,0.82}

\DeclareMathOperator*{\argmax}{arg\,max}

\usepackage{scalerel,stackengine}
\stackMath
\newcommand\reallywidehat[1]{%
\savestack{\tmpbox}{\stretchto{%
  \scaleto{%
    \scalerel*[\widthof{\ensuremath{#1}}]{\kern-.6pt\bigwedge\kern-.6pt}%
    {\rule[-\textheight/2]{1ex}{\textheight}}
  }{\textheight}%
}{0.5ex}}%
\stackon[1pt]{#1}{\tmpbox}%
}
\begin{document}

\title{Vector autoregression models with \\ skewness and heavy tails}
\author{Sune Karlsson, Stepan Mazur and Hoang Nguyen
\\
Örebro University School of Business}

\date{\today}
\maketitle

\begin{abstract}
With uncertain changes of the economic environment, macroeconomic downturns during recessions and crises can hardly be explained by a Gaussian structural shock. 
There is evidence that the distribution of macroeconomic variables is skewed and heavy tailed. 
In this paper, we contribute to the literature by extending a vector autoregression (VAR) model to account for a more realistic assumption of the multivariate distribution of the macroeconomic variables. 
We propose a general class of generalized hyperbolic skew Student's $t$ distribution with stochastic volatility for the error term in the VAR model that allows us to take into account skewness and heavy tails.  
Tools for Bayesian inference and model selection using a Gibbs sampler are provided. 
In an empirical study, we present evidence of skewness and heavy tails for monthly macroeconomic variables. 
The analysis also gives a clear message that skewness should be taken into account for better predictions during recessions and crises.

\textbf{JEL-codes:}  C11, C15, C16, C32, C52


\end{abstract}

\newpage

\section{Introduction} \label{sec:Introduction}
Since the seminal work of \cite{Sims1980}, the vector autoregression (VAR) model has become one of the key macroeconomic models for policy makers and forecasters, see \cite{Karlsson2013}. The utility of the basic VAR model of \citeauthor{Sims1980} has been greatly enhanced by extensions allowing for time varying parameters \citep{Primiceri2005,Cogley2005} and stochastic volatility \citep{Uhlig1997,Clark2011, Clark2015}. These can, however, not fully account for features in the data such as heavy tailed or  skewed distributions.

\cite{Acemoglu2017} gives a theoretical motivation for the non-Gaussian distribution of macroeconomic variables and the presence of heavy tails and asymmetries is well documented in the literature. For example, \cite{Christiano2007} found evidence against Gaussianity by inspecting the skewness and kurtosis properties of residuals from a VAR model and
\cite{Fagiolo2008} find that the distribution of the output growth rates of OECD countries can be approximated by symmetric exponential-power densities with Laplace tails even  after accounting for outliers, autocorrelation and heteroscedasticity. To model the heavy tails \cite{Ni2005} propose a VAR model with a multivariate Student's $t$ distribution, while \cite{Curdia2014} and \cite{Chib2014} impose a similar heavy tailed structural shock in Dynamic Stochastic General Equilibrium (DSGE) models. \cite{Karlsson2020}, on the other hand, propose a general class of multivariate heavy tailed distributions which includes the normal, $t$ and Laplace distributions as well as their mixture for the error term in the VAR model. Stochastic volatility can also lead to a heavy tailed marginal distribution and
\cite{Cross2016}, \cite{Chiu2017} and \cite{Liu2019} show that ignoring the stochastic volatility of the shocks will overestimate the fatness of the tails. 

As noted by, among others, \cite{Curdia2014} the largest shocks occur during recessions, and the skewness of the distribution should be taken into account. Skew-normal and skew-$t$ distributions are common choices for modelling data with skewed distributions. An early application in the VAR literature is \cite{Panagiotelis2008} who proposed the use of a multivariate skew-$t$ distribution. Recently \cite{Liu2019} estimates different asymmetric and heavy tailed distributions for macroeconomic variables, even though the symmetric Student's $t$ distribution is preferred for monthly data.
\cite{Delle2020} model the conditional distribution of GDP using a skew-$t$ distribution with time-varying location, scale and shape parameters. \cite{Carriero2020} apply a VAR model with a common factor volatility in conditional mean and find evidence of skewness in the unemployment rate and the financial indicator. \cite{Carriero2021} account for extreme Covid-19 observations using a VAR model with outlier-augmented stochastic volatility. They show that the model performs on par with a VAR with Student's $t$ distribution. 

In this paper, we contribute to the literature by extending the VAR model to account for more realistic assumptions on the multivariate distribution of the variables. We propose a general class of skewed distributions with heavy tails and stochastic volatility for the error terms in the VAR. In doing so we take the generalized hyperbolic skew Student $t$ distribution as our starting point and we refer to this as the GHSkew-$t$-SV class of VAR models. The GHSkew-$t$-SV distribution can be represented as a normal variance-mean mixture and lends itself to straightforward Bayesian inference using a Gibbs sampler with a few Metropolis-Hastings steps. Model comparison and marginal likelihood calculations can be done using the cross entropy method of \cite{Chan2018}.
 
In an application to monthly US macro data we compare the in-sample and out of sample forecast performance of 14 VAR models with different assumptions on the tail distribution and stochastic volatility. 
We find strong support for the VAR models with skewness and heavy tails. Stochastic volatility, heavy tails and skewness all contribute to the in-sample fit. 
In general, the VAR model with stochastic volatility improves the point and density out-of-sample forecasts. Furthermore, allowing for heavy tailed distributions enhances the out-of-sample forecast which is in agreement with current findings in the literature, see \cite{Chiu2017} and \cite{Liu2019}. An interesting finding is that ignoring the stochastic volatility of the error terms not only overestimates the fatness of the tail distribution but also underestimates the skewness. We recommend that skewness as well as heavy tails should be taken into account for better predictions and in-sample fit. 


The rest of the paper is organized as follows. Section 2 introduces the GHSkew-$t$-SV models. Section 3 presents the Bayesian algorithm for inference and the cross entropy methods to calculate the marginal likelihood. Section 4 illustrates some empirical results with monthly macroeconomic data. Section 5 provides some international evidence with the GHSkew-$t$-SV VAR model. 
Finally, conclusions are reached in Section 6.

\section{VAR Models} \label{sec:Model}

We consider different specifications of the skewness and heavy tails in the distribution of the error terms in the VAR model. 
We also allow for stochastic volatility
in addition to the skewness and heavy tails in the distribution of the error terms.
Starting with the Gaussian VAR model, we extend the Gaussian structural shock to the generalized hyperbolic skew Student's $t$ distribution using the fact that generalized hyperbolic skew Student's $t$ distribution can be obtained as a variance-mean mixture of the Gaussian distribution.

\subsection{A Gaussian VAR Model} \label{sec:GaussModel}

The Gaussian VAR model with stochastic volatility (Gaussian-SV) is given by
\begin{equation}\label{VAR}
\begin{aligned}
\mathbf y_{t} & = \mathbf c + \mathbf B_1 \mathbf y_{t-1} + \ldots + \mathbf B_p \mathbf y_{t-p} + \mathbf A^{-1} \mathbf H_t^{1/2} \boldsymbol \epsilon_t, \ \ \ t=1,\dots,T,\\
\end{aligned}
\end{equation}
where $\mathbf y_{t}$ is a $k$-dimensional vector of endogenous variables; 
$\mathbf c$ is a $k$-dimensional vector of constants; 
$\mathbf B_j$ is a $k \times k$ variate matrix of regression coefficients with $j = 1, \ldots, p$;
$\mathbf A$ is a  $k \times k$ lower triangular matrix with ones on the diagonal that describes the contemporaneous interaction of the endogenous variables;  
$\mathbf H_{t}$ is a $k \times k$ diagonal matrix that captures the heteroskedastic volatility; 
$\boldsymbol {\epsilon}_t$ is a $k$-dimensional vector of error terms that follows a multivariate Gaussian distribution with zero mean vector and identity covariance matrix, i.e.  $\boldsymbol {\epsilon}_t \sim \mathcal N_k(\mathbf 0,\mathbf I)$. 
We assume that the heteroskedastic volatility follows a random walk for $\mathbf H_{t} = diag (h_{1t}, \ldots, h_{kt})$ with
\begin{equation} \label{SVlog}
\begin{aligned}
\log h_{it} = \log h_{it-1} + \sigma_i \eta_{it}, \ \ \ i = 1, \ldots, k,
\end{aligned}
\end{equation}
where $\eta_{it} \sim \mathcal N(0,1)$. 
The Gaussian VAR model without stochastic volatility can be obtained by fixing zero values of $\boldsymbol \sigma^2 = (\sigma_1^2, \ldots, \sigma^2_k)'$ and assuming that $\log h_{it} = \log h_{i0} $ for $i = 1, \ldots, k$ and $t = 1, \ldots, T$.

For notational ease we rewrite the Gaussian VAR model with stochastic volatility in (\ref{VAR}) as 
\begin{equation}\label{VAR2}
\begin{aligned}
\mathbf y_{t} & = \mathbf B \mathbf x_{t} + \mathbf u_{t}, \\
\end{aligned}
\end{equation}
where $\mathbf B = (\mathbf c,  \mathbf B_1,  \ldots,  \mathbf B_{p})$ is a $k \times (1+kp)$ variate matrix, $\mathbf x_t = (1, \mathbf y_{t-1}^{'}, \ldots, \mathbf y_{t-p}^{'} )^{'}$ is $(1+kp)$-dimensional vector and $\mathbf u_{t} = \mathbf A^{-1} \mathbf H_t^{1/2} \boldsymbol \epsilon_t $ is a $k$-dimensional vector of heteroskedastic shocks associated with the VAR equations. 

\subsection{Skew Multivariate \texorpdfstring{$t$}{t} Distributions}

The $t$ distribution has been extended in several different ways to allow for skewness and asymmetric behaviour. Among these, \cite{Ferreira2007} propose a multivariate skew-$t$ distribution via an affine linear transformation of independent skew-$t$ variables while \cite{Sahu2003} use a hidden truncation model to construct a multivariate skew-$t$ distribution where the heavy tail behavior is captured by only one parameter.

We will, however, take the multivariate generalized hyperbolic skew Student's $t$ distribution as our starting point. It is commonly used as it is a general class of distribution which nests the Gaussian distribution and the Student's $t$ distribution as special cases,  see \cite{Mcneil2015}. In addition, \cite{Aas2006} showed that the generalized hyperbolic skew Student's $t$ distribution has exponential/polynomial tail behaviors which can handle substantial skewness in comparison to other skew Student's $t$ distributions. As the generalized hyperbolic skew Student's $t$ distribution can be written in term of a Gaussian variance-mean mixture, it is also straightforward to extend the Gibbs sampler for the Gaussian VAR model with stochastic volatility to accommodate heavy tails and skewness. 

In the next sections, we extend the distribution of the error term $\mathbf u_{t}$ to more flexible multivariate distributions using two approaches. In the first approach, we rewrite the reduced form VAR into a structural VAR and allow for skewness and heavy tails in the orthogonal shocks. In the second approach, we consider the skewness and heavy tails for each marginal distribution directly by assuming a Gaussian variance-mean mixture for the reduced form errors.

\subsection{An Orthogonal Skew-\texorpdfstring{$t$}{Lg} VAR Model} \label{sec:OSkewModel}

We account for the heavy tailed and asymmetric heteroskedastic shocks in the orthogonal residuals  $\mathbf A \mathbf u_{t}$ in the VAR structural form by letting
\begin{equation} \label{OSkewNoise}
\begin{aligned}
\mathbf A (\mathbf y_{t} - \mathbf B \mathbf x_{t})  = \mathbf A \mathbf u_{t} = \mathbf W_t \boldsymbol \gamma + \mathbf W_t^{1/2} \mathbf H_t^{1/2} \boldsymbol \epsilon_t,
\end{aligned}
\end{equation}
where $\boldsymbol \gamma = (\gamma_{1}, \ldots, \gamma_{k})^{'}$ is a $k$-dimensional vector of the skewness parameters, the mixing matrix $\mathbf W_{t} = diag(\xi_{1t}, \ldots, \xi_{kt})$ is a $k \times k$ diagonal matrix with $\xi_{it}$ that follows inverse Gamma distribution with shape parameter $\nu_i/2$ and rate parameter $\nu_i/2$, i.e. $\xi_{it} \sim \mathcal {IG} (\frac{\nu_i}{2}, \frac{\nu_i}{2})$, and $\boldsymbol \nu = ( \nu_1, \ldots, \nu_k )^{'}$ is a  $k$-dimensional vector that consists of the degrees of freedom; 
$\mathbf W_t$ and $\boldsymbol \epsilon _t$ are independently distributed.
Equation (\ref{OSkewNoise}) represents the marginal distribution of the orthogonal shock $\mathbf A \mathbf u_{t}$ as a vector of independent univariate generalized hyperbolic skew Student's $t$ distributions (OST). By setting $\boldsymbol\gamma$ to zero we obtain the symmetric and orthogonal $t$ distribution (OT) used by \cite{Curdia2014}, \cite{Clark2015} and \cite{Chiu2017}. As usual, letting the degree of freedom $\nu_i \rightarrow \infty$ for $i = 1, \ldots, p$, the OT VAR model becomes a Gaussian VAR.

Given the mixing matrix $\mathbf W_{t}$, it holds that 
$$
\mathbf u_t| \mathbf W_t 
\sim 
\mathcal N_k \left( \boldsymbol \mu_t = \mathbf A^{-1} \mathbf W_{t} \boldsymbol\gamma, \boldsymbol \Sigma_t = \mathbf A^{-1} \mathbf W_t^{1/2} \mathbf H_t \mathbf W_t^{1/2}\mathbf A^{-1'}  \right).
$$
\cite{Chiu2017} interprets the mixing matrix $\mathbf W_t$ as capturing the high-frequency shocks in mean and volatility while the stochastic volatility accounts for the low-frequency shocks. 
The data will determine whether the extreme time variation comes from the volatility shift or from the idiosyncratic heavy tail shocks.

\subsection{A Multi-Skew-\texorpdfstring{$t$}{Lg} VAR Model} \label{sec:mSkewModel}

The OST VAR builds the distribution of the error terms from the ground up in terms of the structural form innovations. This makes for a straightforward structural interpretation but also means that the model is sensitive to the identifying assumptions, in this case the triangular structure of $\mathbf A$ and the ordering of the variables. To overcome this we can model the reduced form errors directly as a correlated vector of univariate skew $t$ distributions.
We propose a class of multi skew-$t$ (MST) VAR models by assuming that the residuals $\mathbf u_{t}$ are given by
\begin{equation} \label{mSkewNoise}
\begin{aligned}
\mathbf u_{t} = \mathbf W_t \boldsymbol \gamma + \mathbf W_t^{1/2} \mathbf A^{-1} \mathbf H_t^{1/2} \boldsymbol \epsilon_t.
\end{aligned}
\end{equation}
The MST VAR model imposes the skewness and heavy tails directly on the reduced form error in each VAR equation rather than in the idiosyncratic shock. 
Hence, the marginal distribution of the endogenous variable $y_{it}$ is generalized hyperbolic skew Student's $t$ for $i = 1, \ldots, k$ and $t = 1, \ldots, T$.
On the other hand, the OST VAR model considers the reduced form error as a linear combination of orthogonal skew-$t$ shocks. Restricting the mixing variables to be equal for the different equations, $\xi_{1t} = \ldots = \xi_{kt}$, induces a common tail behaviour and the marginal distribution of $\mathbf u_{t}$ is a multivariate generalized hyperbolic skew Student $t$ (Skew-$t$) distribution  \citep{Mcneil2015}. 
Additionally, setting $\gamma_{1} = \ldots = \gamma_{k} = 0$ results in a multivariate Student $t$ (Student-$t$) distribution.

The defining feature of the multi skew-$t$ in \eqref{mSkewNoise} and what differentiates it from the usual (skew) multivariate $t$ is the equation specific mixing variables $\xi_{it}$. There is thus no commonality in the tail behaviour of the equations.

Also, if $\nu_i \rightarrow \infty$, the MT model becomes a Gaussian VAR with stochastic volatility in spirit of \cite{Cogley2005} and \cite{Primiceri2005}.

In the next section, we illustrate the Bayesian inference and model selection criteria for different specifications of VAR models with/without stochastic volatility and different assumptions on the distribution of the error terms such as Gaussian, Student-$t$, Skew-$t$, orthogonal Student's $t$ (OT), multi Student's $t$ (MT), orthogonal skew Student's $t$ (OST), multi skew Student's $t$ (MST).

\section{Bayesian Inference}

In this section, we discuss the prior distributions for the parameters in the MST-SV VAR models and the corresponding general Gibbs sampler scheme. 
Inference procedures for other model variations are described when needed. 
We also show how to compute the marginal likelihood based on the cross entropy method proposed by \cite{Chan2018}.

\subsection{Prior Distribution}

Denote the set of the MST-SV VAR model parameters and latent variables by \newline
$\boldsymbol \theta = \{ vec(\mathbf B)^{'}, \mathbf a^{'}, \boldsymbol \gamma^{'}, \boldsymbol \nu^{'}, \boldsymbol \sigma^{2'}, \mathbf \xi_{1:K,1:T}^{'}, \mathbf h_{1:K,0:T}^{'}\}^{'}$, where $\mathbf a = (a_{2,1}, a_{3,1}, a_{3,2}, \ldots, a_{k,k-1})'$ is the set of elements of the lower triangular matrix $\mathbf A$ and $\mathbf h_{1:K,0}$ is the vector of initial values for the stochastic volatilities. 
We employ the Minnesota priors for the prior distributions of $\mathbf B$ with the overall shrinkage $l_1 = 0.2$ and the cross-variable shrinkage $l_2 = 0.5$, see \cite{Koop2010}, and vague prior distributions for other parameters. 
In details, the Minnesota-type priors assume a Gaussian prior for $vec(\mathbf B)$, i.e. $vec(\mathbf B) \sim \mathcal N ( \mathbf b_{0}, \mathbf V_{\mathbf b_{0}})$, 
that shrinks the regression coefficients towards univariate random walks with a tighter prior around zero for longer lags. 
The prior for $\mathbf a$ is also Gaussian, $\mathbf a \sim \mathcal N_{0.5k(k-1)} (0, 10  \mathbf I)$, which implies a weak assumption of no interaction among endogenous variables.
The parameters which account for the heavy tails are endowed with Gamma priors, $\nu_i \sim \mathcal G (2,0.1)$ for $i = 1, \ldots, k$ and the skewness parameters are given a normal prior, $\boldsymbol \gamma \sim \mathcal N_k ( \mathbf 0, \mathbf I)$. That is the prior mean of the degrees of freedom of the $t$-distribution is 20 and the skewness has zero prior mean. 
Finally, the prior for the variance of the shock to the volatility is $\sigma_i^2 \sim \mathcal{G} ( \frac{1}{2}, \frac{1}{2V_{\sigma}})$ which is equivalent to $\pm \sqrt{\sigma_i^2} \sim \mathcal N (0,V_{\sigma})$ see \cite{Kastner2014}, this prior is less influential in comparison to the conjugated inverse Gamma prior especially when the true value is small. In all cases of VAR model with and without stochastic volatility $\log h_{i0} \sim \mathcal N \left( \log \hat{\Sigma}_{i,OLS}, 4 \right)$ where $\hat{\Sigma}_{i,OLS}$ is the estimated variance of the AR(p) model using the ordinary least square method, see \cite{Clark2015}.

\subsection{Estimation Procedure}

%

Given the latent variables $\mathbf \mathbf \xi_{1:K,1:T}$ and the skewness parameters $\boldsymbol \gamma$, the conditional posterior distributions of the remaining parameters in the MST-SV VAR model are similar to those in the Gaussian-SV VAR model. 
Hence, the MST model can be estimated using a seven-step Metropolis-within-Gibbs Markov chain Monte Carlo (MCMC) algorithm. 
Let's $\boldsymbol \Psi$ be a set of conditional parameters except the one that we sample from.

\begin{enumerate}

\item
In order to sample from $\pi (vec(\mathbf B) | \boldsymbol \Psi)$, Equation (\ref{VAR2}) can be rewritten as a multivariate linear regression,
\begin{equation*} 
\begin{aligned}
\mathbf y_{t} -  \mathbf W_t \boldsymbol \gamma & = \mathbf B \mathbf x_{t} +  \boldsymbol \Sigma_t^{1/2} \boldsymbol \epsilon_t,
\end{aligned}
\end{equation*}
where $\boldsymbol \Sigma_t^{1/2} = \mathbf W_t^{1/2} \mathbf A^{-1} \mathbf H_t^{1/2}$. Then the conditional posterior distribution of $vec(\mathbf B)$ is a conjugate Gaussian distribution
$$ \pi (vec(\mathbf B) | \boldsymbol \Psi) \sim \mathcal N (\mathbf b^*, \mathbf V_{\mathbf b}^*),$$ 
where
\begin{equation*} 
\begin{aligned}
\mathbf V_{\mathbf b}^{*-1} =& \mathbf V_{\mathbf b_{0}}^{-1} + \sum_{t=1}^{T} \mathbf x_t \mathbf x^{'}_t \otimes \boldsymbol \Sigma_t^{-1},  \\
\mathbf b^* =& \mathbf V_{\mathbf b}^{*} \left[ \mathbf V_{\mathbf b_{0}}^{-1} \mathbf b_0 + \sum_{t=1}^{T} ( \mathbf x_t \otimes \boldsymbol \Sigma_t^{-1} (\mathbf y_t - \mathbf W_t \boldsymbol \gamma))   \right].
\end{aligned}
\end{equation*}

\item 
In order to sample from $\pi (\boldsymbol \gamma | \boldsymbol \Psi)$, we consider equation (\ref{VAR2}) as a multivariate linear regression, 
\begin{equation*} 
\begin{aligned}
\mathbf y_{t} -  \mathbf B \mathbf x_{t} & =  \mathbf W_t \boldsymbol \gamma +  \boldsymbol \Sigma_t^{1/2} \boldsymbol \epsilon_t.
\end{aligned}
\end{equation*}
Hence, the conditional posterior distribution of $\boldsymbol \gamma$ is a conjugate Gaussian distribution 
$$\pi (\boldsymbol\gamma | \boldsymbol \Psi) \sim \mathcal N_k (\boldsymbol \gamma^*, \mathbf V_{\boldsymbol \gamma}^*),$$ 
where
\begin{equation*} 
\begin{aligned}
\mathbf V_{\boldsymbol \gamma}^{*-1} =& \mathbf V_{\boldsymbol \gamma}^{-1} + \sum_{t=1}^{T} \mathbf W^{'}_t \boldsymbol \Sigma_t^{-1} \mathbf W_t ,  \\
\boldsymbol \gamma^* =&  \mathbf V_{\boldsymbol\gamma}^{*} \left(  \sum_{t=1}^{T} \mathbf W^{'}_t \boldsymbol \Sigma_t^{-1} \left(\mathbf y_{t} -  \mathbf B \mathbf x_{t} \right) \right).
\end{aligned}
\end{equation*}

\item 
In order to sample from $\pi ( \mathbf a| \boldsymbol \Psi)$, we follow \cite{Cogley2005} and use that (\ref{mSkewNoise}) is a triangular model for the reduced form residuals,
\begin{equation*} 
\begin{aligned}
\mathbf A \tilde{\mathbf u}_t  & =  \mathbf H_t^{1/2} \boldsymbol \epsilon_t,
\end{aligned}
\end{equation*}
where $\tilde{\mathbf u}_t = \mathbf  W_t^{-1/2} (\mathbf y_{t} -  \mathbf B \mathbf x_{t} - \mathbf W_t \boldsymbol \gamma)$. 
This reduces to a system of linear equations with equation $i$ that has $\tilde{u}_{it}$ as a dependent variable and $- \tilde{u}_{jt}$ as independent variables with coefficients $a_{ij}$ for $i = 2, \ldots, k$ and $j = 1, \ldots, i-1$. 
By multiplying both sides of the equations with $h_{it}^{- 1/2}$, we can eliminate the effect of heteroscedasticity. 
Then, draws from the conditional posterior of $a_{ij}$ can be taken equation by equation using the conditionally Gaussian posterior distribution \citep{Cogley2005}.

\item 
In order to sample from $\pi (\mathbf h_{1:K,0:T}^{'} | \boldsymbol \Psi)$, we follow  \cite{Kim1998,Primiceri2005, Del2015}. 
Let $\tilde{\tilde{\mathbf u}}_{t} = \mathbf A \tilde{\mathbf u}_{t}$, for each series $i = 1,\ldots, k$, we have that $\log \tilde{\tilde{u}}_{it}^2 = \log h_{it} + \log \epsilon_t^2 $.
\cite{Kim1998} approximated the $\chi^2$ distribution  of $\epsilon_t^2$ using a mixture of 7 Gaussian components. 
Then using forward filter backward smoothing algorithm in \cite{Carter1994}, we sample $\log h_{it}$ from its smoothing Gaussian distribution.


\item 
In order to sample from $\pi (\boldsymbol \sigma^2 | \boldsymbol \Psi)$, Equation (\ref{SVlog}) describes a random walk in the logarithm of the volatility. 
The conditional posterior $\pi ( \sigma^2_i |\boldsymbol \Psi)$ is 
$$
\pi (\sigma^2_i | \boldsymbol \Psi) 
\propto \left( \sigma_i^2 \right)^{-\frac{T}{2}} \exp \left(- \frac{ \sum_{t = 1}^{T} (\log h_{it} - \log h_{it-1})^2}{ 2 \sigma_i^2 } \right) \left( \sigma_i^2 \right)^{-\frac{1}{2}} \exp \left(- \frac{\sigma_i^2}{2 V_{\sigma}} \right).$$ 
 We apply the independent Metropolis-Hastings algorithm to draw $\sigma_i^{2(*)} \sim \mathcal{IG}(\alpha_i, \beta_i)$ where $\alpha_i = \frac{1}{2}T$ and $\beta_i = \frac{1}{2}\left( \displaystyle\sum_{t = 1}^T (\log h_{i,t} - \log h_{i,t-1} )^2\right)$ and accept with the probability 
 \begin{equation*}
\begin{aligned}
& \min \left\{ 1,  \frac{\sigma_i^{(*)} }{\sigma_i} \exp \left( \frac{\sigma_i^{2} - \sigma_i^{2(*)}}{2V_{\sigma}} \right) \right\} 
\end{aligned}
\end{equation*}

\item 
In order to sample from $\pi (\nu_i | \boldsymbol \Psi) \propto \mathcal G(\nu_i; 2, 0.1) \displaystyle\prod_{t = 1}^T \mathcal{IG} \left(\xi_{it}; \frac{\nu_i}{2} , \frac{\nu_i}{2}\right) $ for $i=1,\ldots, k$,  we use an adaptive random walk Metropolis-Hastings algorithm to accept/reject the draw $\nu_i^{(*)} = \nu_i + \eta_i \exp(c_i)$, where $\eta_i \sim \mathcal N(0,1)$ and the adaptive variance $c_i$ is adjusted automatically such that the acceptance rate is around 0.25 \citep{Roberts2009}.

\item 
In order to sample $\pi (\xi_{1:K,t} | \boldsymbol \Psi)$ for $t = 1,\ldots, T$, we apply the independent Metropolis-Hastings algorithm to draw $\xi_{it}^{(*)} \sim \mathcal{IG}(\alpha_{it}, \beta_{it})$ for $i = 1, \ldots, K$ and accept with the probability 
\begin{equation*}
\begin{aligned}
& \min \left\{ 1,  \frac{\pi (\mathbf W_{t}^{(*)} | \boldsymbol \Psi) \displaystyle\prod_{i = 1}^k  \mathcal{IG}(\xi_{it}; \alpha_{it}, \beta_{it})}{\pi (\mathbf W_{t} | \boldsymbol \Psi) \displaystyle\prod_{i = 1}^k  \mathcal{IG} (\xi_{it}^{(*)}; \alpha_{it}, \beta_{it})} \right\} 
\end{aligned}
\end{equation*}
where  
\begin{eqnarray*}
\pi (\mathbf W_{t} | \boldsymbol \Psi) 
\propto 
\displaystyle\prod_{i = 1}^k \xi_{it}^{-1/2} \exp \left( -\frac{1}{2} (\mathbf y_t-\mathbf B \mathbf x_t - \mathbf W_t \boldsymbol \gamma)^{'} \boldsymbol \Sigma_t^{-1} (\mathbf y_t- \mathbf B \mathbf x_t - \mathbf W_t \boldsymbol \gamma) \right) 
\mathcal{IG}\left(\xi_{it}; \frac{\nu_i}{2}, \frac{\nu_i}{2}\right).
\end{eqnarray*}
The proposal distribution $\mathcal {IG} (\alpha_{it}, \beta_{it})$ is taken from \cite{Chiu2017} with $\alpha_{it} = \frac{c}{2} (\nu_i+1)$ and $\beta_{it} = \frac{c}{2} \left(\nu_i+ \frac{\tilde{\tilde{u}}_{it}^2}{ h_{it}} \right)$ where the constant $c = 0.75$ is adjusted so that the acceptance rate range from 20\% to 80\%. 
In OST-SV model, we can sample the conditional posterior $\pi (\xi_{i,t} | \boldsymbol \Psi)$ independently for $i = 1,\ldots,K$. 
We find the mode and inverse Hessian at the mode of the full conditional log-posterior of $\pi (\xi_{it} | \boldsymbol \Psi)$ and draw a proposal value  $\log(\xi_{it})$ from a $t$ distribution with four degrees of freedom with mean equal to the mode and scale equal to the inverse Hessian at the mode, see \cite{Creal2015} and \cite{Nguyen2019}. 

\end{enumerate}

\subsection{Model Selection}
\label{sec: model selection}

The marginal likelihoods of the GHSkew-$t$-SV VAR models require the high-dimensional integration
\begin{equation} \label{Ml}
\begin{aligned}
p(\mathbf y_{1:T}) = \int p(\mathbf y_{1:T}| \boldsymbol \theta) \pi (\boldsymbol \theta) d\boldsymbol \theta .
\end{aligned}
\end{equation}
Following the adaptive importance sampling approach of \cite{Chan2018}, we divide the model parameters into two groups with the static parameters $\boldsymbol \theta_1 = \{ vec(\mathbf B)^{'}, \mathbf a^{'}, \boldsymbol \gamma^{'}, \boldsymbol \nu^{'}, \boldsymbol \sigma^{2'}, \mathbf h_{0}^{'}\}^{'}$ and the latent states $\boldsymbol \theta_2 = \{ \mathbf \xi_{1:K,1:T}^{'}, \mathbf h_{1:K,1:T}^{'} \}^{'}$. 
We first use the cross-entropy methods to a the proposal distribution for $\boldsymbol \theta_1$,  $f(\boldsymbol \theta_1)$, from the posterior samples. 
Then, the integrated likelihood $p(y_{1:T}|\boldsymbol \theta_1)$ is calculated using an inner importance sampling loop based on a sparse matrix representation. 
Algorithm 1 summarizes the marginal likelihood calculation using the adaptive importance sampling approach of \cite{Chan2018}.

\textbf{Algorithm 1.} (Marginal likelihood estimation via the cross-entropy method)
\begin{enumerate}
\item 
Obtain the posterior samples $\boldsymbol \theta_1^{(1)}, \ldots, \boldsymbol \theta_1^{(R)}$ from the posterior density $\pi(\boldsymbol \theta|\mathbf y_{1:T})$.
\item 
Consider the parametric family $f(\boldsymbol \theta_1;\lambda)$ parameterized by parameter $\lambda$ such that 
$$\lambda^{*} = \argmax_{\lambda} \frac{1}{R} \sum_{r=1}^{R} \log f( \boldsymbol \theta_1^{(r)}|\lambda)$$
\item 
Obtain new samples $\boldsymbol \theta_1^{(1)}, \ldots, \boldsymbol \theta_1^{(N)}$ from $f(\boldsymbol \theta_1;\lambda^{*})$. 
For each new value $ \boldsymbol \theta_1^{(n)}$, we show in the appendix how to obtain the integrated likelihood $\hat{p}(\mathbf y_{1:T}|\theta_1^{(n)})$. Then the marginal likelihood is calculated via importance sampling
\begin{equation*}
\begin{aligned}
\hat{p}_{IS}(\mathbf y_{1:T}) &=  \frac{1}{N} \sum_{n = 1}^{N} \frac{\hat{p}(\mathbf y_{1:T}|\boldsymbol \theta_1^{(n)}) p(\boldsymbol \theta_1^{(n)})}{f(\boldsymbol \theta_1^{(n)} | \lambda^{*})}.
\end{aligned}
\end{equation*}
\end{enumerate}
The number of samples $N$ is chosen such that the variance of the estimated quantity using important sampling is less than one. 
The parametric families of $f(\boldsymbol \theta_1)$ are the multivariate Gaussian distribution for $(\mathbf B,\mathbf a,\boldsymbol \gamma_{1:k})$, the independent Gamma distribution for $\nu_{1:k}$ and independent inverse Gamma distribution for $\sigma_{1:k}^2$ and $h_{1:k,0}^2$.

\subsection{Forecast metrics} \label{Metrics}

In addition to the in-sample model comparison we also assess the forecasting performance of the different specifications of the error distribution in a recursive out of sample forecasting exercise.
We compare the forecast accuracy using the mean square forecast error (MSFE) for the point forecast, the log predictive density (LP), and the continuous rank probability score (CRPS) of the posterior predictive distribution for the density forecast. 

Let $T_0$ be the last observation in the first estimation sample and $T_1$ the last observation on variable $i$. The MSFE of variable $i$ at $h$ step ahead, for $h = 1,  \ldots, H$,  is then obtained as,
\begin{equation*}
\begin{aligned}
\mathrm{MSFE}_{i,h} &=  \frac{1}{T_1-T_0-h+1} \sum_{t = T_0}^{T_1-h} \left( \bar{y}_{i,t+h|t} - y_{i,t+h}^{o} \right)^{2},
\end{aligned}
\end{equation*}
where $\bar{y}_{i,t+h|t}$ is the mean of the posterior predictive samples using all data up to time $t$ and  $y_{i,t+h}^{o}$ is the observed outcome of variable $i$ at $h$ steps ahead. The model with a smaller MSFE is preferred.

The LP of the posterior predictive distribution is computed as,
\begin{equation*}
\begin{aligned}
\mathrm{LP}_{i,h} &=  \frac{1}{T_1-T_0-h+1} \sum_{t = T_0}^{T_1-h} \Bigg[ \log p(y_{i,t+h}^{o} | \mathbf y_{1:t})  \Bigg]
\\
&= \frac{1}{T_1-T_0-h+1} \sum_{t = T_0}^{T_1-h} \Bigg[ \log \displaystyle\int_{\boldsymbol \theta}  p( y_{i,t+h}^{o} | \boldsymbol \theta,  \mathbf y_{1:t}) p( \boldsymbol \theta|  \mathbf y_{1:t}) d \boldsymbol \theta  \Bigg] \\
\end{aligned}
\end{equation*}
where $p(y_{i,t+h}^{o} | \mathbf y_{1:t})$ is the $h$-step ahead posterior predictive density function evaluated at the realization of the variable.  
Following \cite{Andersson2008}, the LP of the posterior predictive distribution is computed using the Rao-Blackwellization idea which is more stable than the kernel density estimator for extreme observations.
In particular, it is evaluated as, 
\begin{equation*}
\begin{aligned}
\mathrm{LP}_{i,h} &=  \frac{1}{T_1-T_0-h+1} \sum_{t = T_0}^{T_1-h} \Bigg[ \log \sum_{r = 1}^{R} \frac{1}{R} p(y_{i,t+h}^{o} | \boldsymbol \theta^{(r)},  \mathbf y_{1:t})  \Bigg] \\
\end{aligned}
\end{equation*}
where $\boldsymbol \theta^{(1)}, \ldots, \boldsymbol \theta^{(R)}$ are the posterior samples of the VAR model. 
The possibly high dimensional integral over intermediate observations implicit in $p(y_{i,t+h}^{o} | \boldsymbol \theta^{(r)},  \mathbf y_{1:t})$ can be approximated by the Monte Carlo approach. For each sample from the posterior we simulate a new path $\mathbf y_{(t+1):(t+h-1)|t}^{(r)}$ using the data generating process for the model and calculate $p(\mathbf y_{i,t+h|t} | \boldsymbol \theta^{(r)},  \mathbf y_{1:t}, \mathbf y_{(t+1):(t+h-1)|t}^{(r)})$.  A higher LP value indicates a better density forecasting performance of the model.

The continuous rank probability score (CRPS) is also commonly used to rank the density forecasts. CRPS is obtained as the quadratic
difference between the predictive cumulative distribution function and the empirical distribution of the variable \citep{Gneiting2007}.
As \cite{Clark2015} noted the CRPS is less sensitive to outliers than the LP and rewards more for values of the predictive density that are close to the outcome. 
\begin{equation*}
\begin{aligned}
CRPS_{i,h} &= \frac{1}{T_1-T_0-h+1} \sum_{t = T_0}^{T_1-h} \Bigg[ E_{f} \left| y_{i,t+h|t} - y_{i,t+h}^{o} \right| - 0.5 E_{f} \left| y_{i,t+h|t} - y_{i,t+h|t}^{'} \right|\Bigg] ,
\end{aligned}
\end{equation*}
where $f$ is the predictive density of the variable $y_{i,t+h|t}$, and $(y_{i,t+h|t}, y_{i,t+h|t}^{'}) $ are independent random draws from the predictive density $f$. 
We apply the Monte Carlo method to simulate 10,000 draws from the predictive density $f$ and compute the expectation.

\section{Empirical Illustration}

We estimate a four-variable VAR with industrial production, inflation rate, unemployment rate, Chicago board options exchange's volatility index (VIX) to illustrate the performance and empirical relevance of the different specifications of the error distribution. We use monthly data for the period 01/1970 to 12/2019 from the Federal Reserve Bank of St. Louis, see \cite{McCracken2016}. Industrial production is included as a growth rate (first difference of the logarithm of the index), the inflation rate is calculated as the first difference of the log of the CPI and the logarithm of the VIX is used. The variables enter with $p=4$ lags.

We compare 14 different specifications for the error distribution: Gaussian; multivariate Student's $t$ and multivariate skew-$t$; orthogonal $t$ (OT) and orthogonal skew-$t$ (OST); multi $t$ (MT) and multi skew-$t$ (MST). All with and without stochastic volatility.

We first estimate the 14 VAR models with and without stochastic volatility using the in-sample dataset. Then 
we perform an out-of-sample forecasting exercise to measure the forecast accuracy of each VAR model.

\subsection{In-sample Analysis}

\begin{figure}[!htbp] 
    \includegraphics[width=1\textwidth]{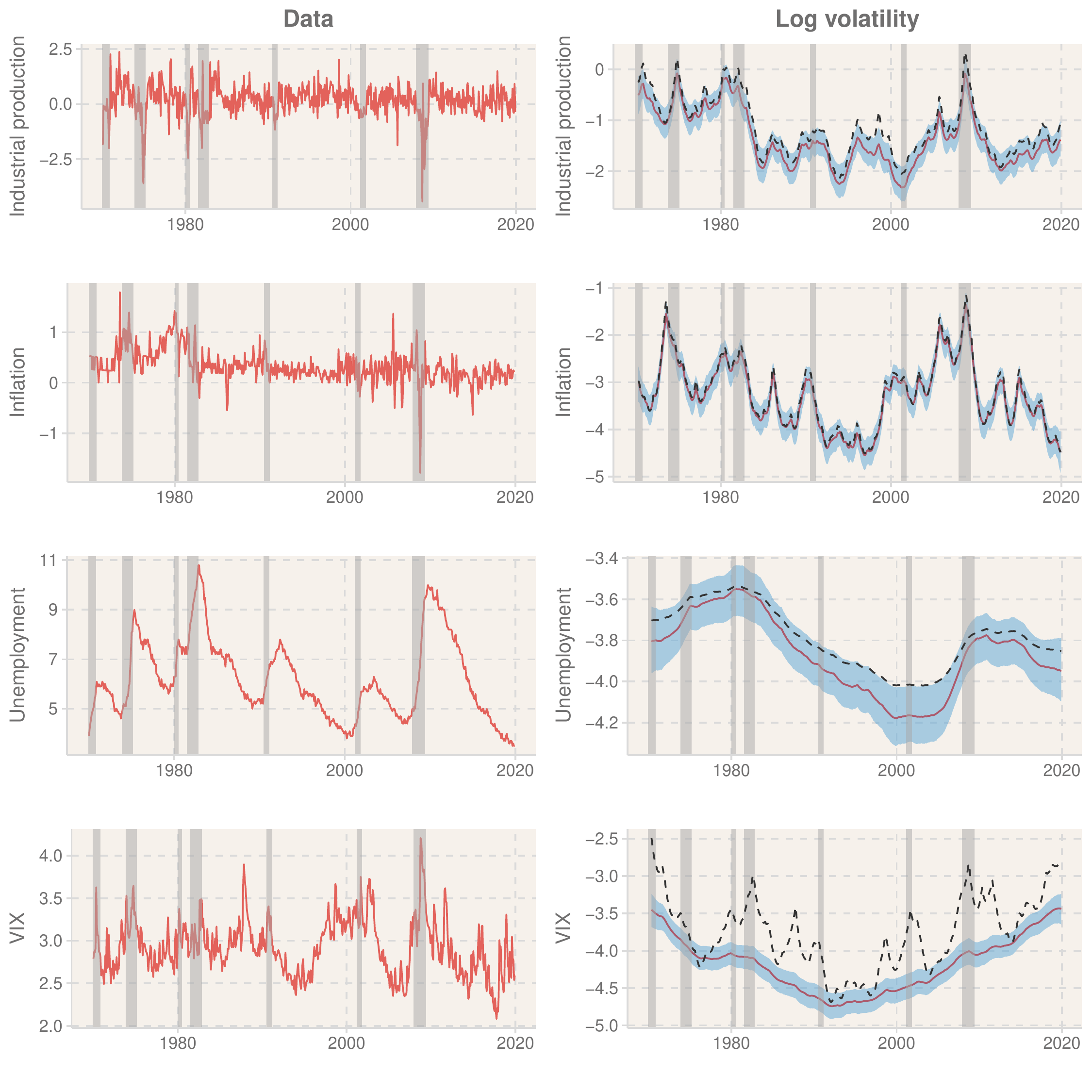}
    \caption{Data and estimated volatilities.}
    \label{fig:samplesbifcop1}
    \caption*{{\scriptsize The figures on the left-hand side show the variables while the figures on the right-hand side draw the estimated mean log volatility of the OST-SV model using a solid line (red) with their 50\% credible interval. The dashed line shows the estimated mean log volatility from the Gaussian-SV model. The shaded areas highlight the recession periods based on the NBER indicators. } }
\end{figure}

The left-hand side of Figure \ref{fig:samplesbifcop1} shows the growth rate of industrial production, inflation rate, unemployment rate and the VIX. Extreme values of the variables are often observed during recession periods based on the NBER indicators. Industrial production growth decreased by more than 4\% during the financial crisis in 2008, while the unemployment rate peaked at 10\% and the VIX reached as high as 4.2. These unconditional skewed behaviors can be generated by a time-varying variance shock and/or a skewed shock. The right-hand side of Figure \ref{fig:samplesbifcop1} plots the estimated stochastic volatility in the log scale. The volatilities are occasionally higher during recessions which illustrates the relation of the VIX and other macroeconomic variables. We compare the volatilities of the low frequency shocks obtained from the OST-SV model in the solid lines and that of the Gaussian-SV model in the dashed lines.
 Using the Gaussian-SV model, the volatility of macroeconomic variables might be overestimated during the recessions and crises which is in agreement with the finding of \cite{Curdia2014}, \cite{Chiu2017}, among others.

Table \ref{LML} estimates the log marginal likelihood of the VAR models of Section \ref{sec:Model} with and without stochastic volatility. In the class of VAR models without SV, allowing for heavy tails leads to a substantial improvement in the marginal likelihood while the addition of skewness is less useful. Allowing for stochastic volatility leads to a dramatic improvement in the marginal likelihood for all seven specifications. Allowing for heavy tails improves on the Gaussian-SV and the more flexible OST and MST specifications of skewness perform best with a log Bayes factor of 4.8 (MST) and 3.9 (OST) against the third best Student-$t$ specification. It is interesting that skewness plays a more important role in the VAR model with SV than in the VAR model without SV. The flexibility of the tail behaviour in the OST and MST is, however, important as evidenced by the relatively poor performance of the skew-$t$ VAR models where only one mixing variable is used to model the heavy tails. 

\begin{table}[!tbp] 
\centering
\caption{Log marginal likelihood for VAR models with and without stochastic volatility} \label{LML}
\resizebox{0.85\textwidth}{!}{%
\begin{threeparttable}
\begin{tabular}{@{}ccrrrrrrr@{}}
\hline
 &  & \multicolumn{1}{c}{Gaussian} & \multicolumn{1}{c}{Student-$t$} & \multicolumn{1}{c}{Skew-$t$} & \multicolumn{1}{c}{OT}  & \multicolumn{1}{c}{MT} & \multicolumn{1}{c}{OST} & \multicolumn{1}{c}{MST} \\ \hline
 & Non SV & -204.8 & -110.6 & -118.8 & -110.5 & -107.8 & -110.1 & -107.2 \\ 
 & SV & -48.1 & -31.1 & -38.2 & -37.5 & -34.3 & -27.2 & -26.3 \\ 
\hline
\end{tabular}
\begin{tablenotes}
      \item {\footnotesize
      We compare the LMLs of 14 VAR models with/without SV. We use the cross entropy methods by \cite{Chan2018} to calculate the LMLs.
      We first sample 20,000 draws from the conditional posterior distributions with 5,000 draws as burn-in. Then, all LMLs estimated using 20,000 draws from the proposal distributions, see details in Section \ref{sec: model selection}.
      }
    \end{tablenotes}
  \end{threeparttable}  
}
\end{table}

\begin{figure}[!htbp] 
    \includegraphics[width=1\textwidth]{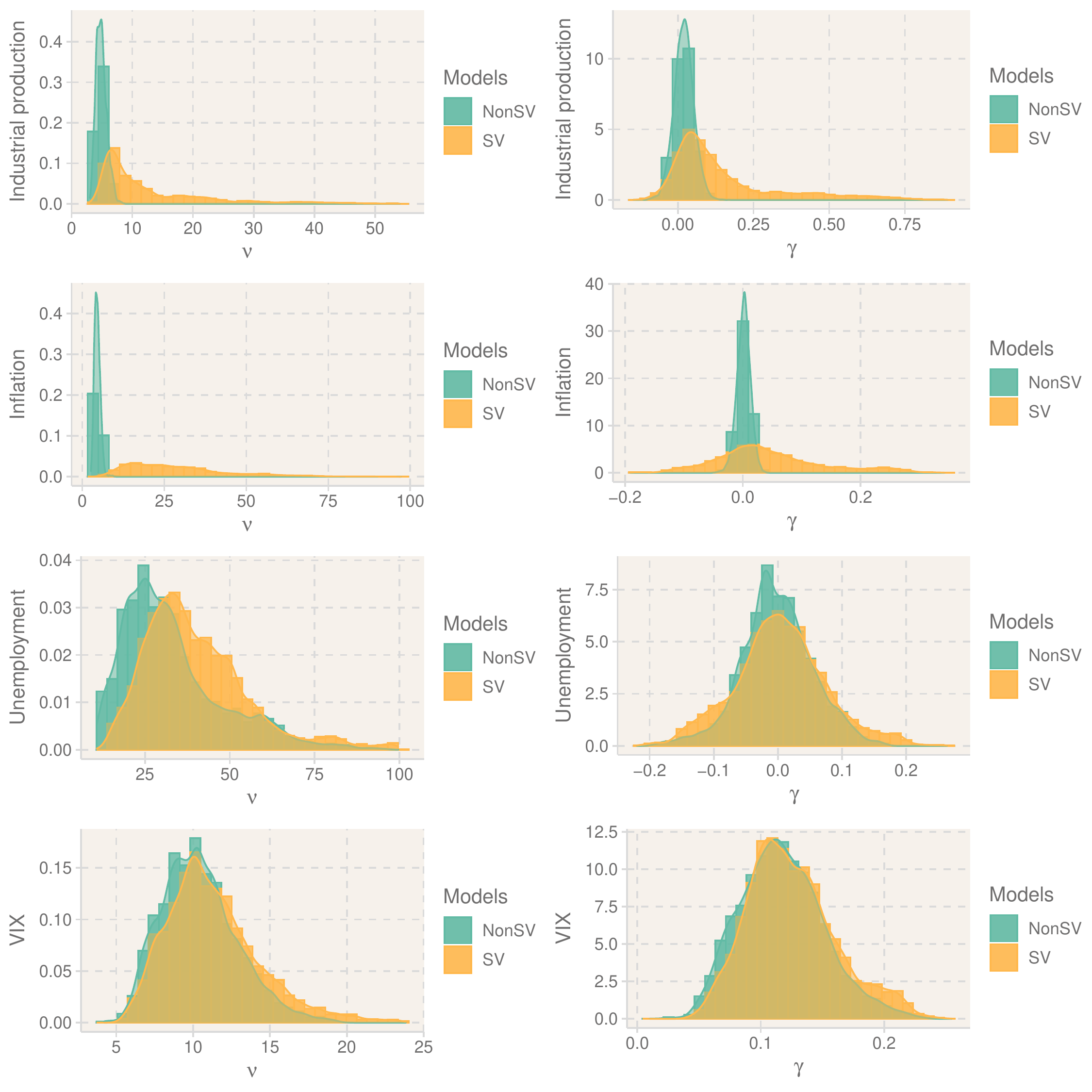}
    \caption{Posterior distribution of the heavy tail (left column) and skewness (right column) parameters of the MST VAR model with and without stochastic volatility.}
    \label{fig:NuSVvsnonSV}
    \caption*{  }
\end{figure}

Next, we take a closer look at the effect of the SV assumption on the skewness and heavy tail parameters in the VAR models.
Figure \ref{fig:NuSVvsnonSV} focuses on the posterior samples of the skewness parameters and the degree of freedom parameters in the MST VAR model with and without stochastic volatility. The left hand side figures show that ignoring the time-varying volatility of the structural shocks overestimates the fatness of the tails. This finding is inline with  \cite{Chiu2017} and \cite{Liu2019}, among others. Moreover, we find that the asymmetric distribution is more important in the VAR models with SV. The magnitudes of the skewness parameters for industrial production growth and inflation rate are higher in the case of SV models while the fatness of the tail distribution are smaller. Appendix \ref{PostApp} also confirms this finding for the OST-SV VAR model.
Hence, even considering that the stochastic volatility has captured the time varying uncertainty, there are still asymmetric structural shocks between recession and moderation periods. 
The heavy tails appear in both industrial production and VIX, however, VIX shows some degree of skewness in the right tail. Although the conditional distributions of inflation rate and unemployment rate are closer to the Gaussian case with SV, there is a possibility that inflation rate is right-skewed. To summarize, when the stochastic volatility assumption reduces the degree of heavy tailed shock, it also increases the level of skewness shock.

\newpage

\subsection{Out-of-sample forecast}

To assess the out-of-sample predictive accuracy of the different specifications, we conduct a recursive forecast exercise using the 01/2017 to 12/2019 period as our evaluation sample. We calculate the MSFE for the point forecasts, and the LP and CRPS for the density forecasts.
 As the VAR models can be nested based on the assumptions on the tail distributions, they are divided into two model groups without and with stochastic volatility for ease of comparison. Using the Gaussian VAR as a benchmark in each group, we test the forecast accuracy using the one-sided \cite{Diebold1995} test where the standard errors of the test statistics are computed with the Newey–West estimator, see the discussion in \cite{Clark2011}. We also compare the Gaussian model with SV to the Gaussian model without SV.

Table \ref{tMSFE} reports the relative improvements in MSFE over the Gaussian VAR models.
Each panel reports the MSFE of each variable using the Gaussian VAR model with (and without) stochastic volatility. The relative improvements over the Gaussian models are computed as the ratio of the MSFE of alternative specifications over the Gaussian models during 2007-2019. 
Entries less than 1 indicate that the given model is better. 
In the non stochastic volatility VAR group, skewness and heavy tail models improve the point forecast of Industrial production growth up to 12 months ahead but are only statistically significant up to 6 months ahead. While the prediction of unemployment rate is improved in the long run and that of inflation is enhanced in the one step ahead forecast. In the stochastic volatility VAR group, the advantage of skewness and heavy tail models over Gaussian model diminishes. For unemployment there is an improvement overall when allowing for heavy tails and or skewness, significantly so for the one month ahead forecasts.

\begin{table}[!htbp] 
\centering
\caption{Relative improvements in MSFE over the Gaussian VAR models} \label{tMSFE}
\resizebox{1\textwidth}{!}{%
\begin{threeparttable}
\begin{tabular}{*{1}{p{2.5cm}} *{8}{p{1.65cm}}}
  \hline
 & 1M & 3M & 6M & 12M & 1M & 3M & 6M & 12M \\   \hline
&  \multicolumn{4}{c}{(a) Industrial Production} & \multicolumn{4}{c}{(c) Unemployment rate} \\  \hline
\rowcolor{light-gray} Gaussian & 0.469 & 0.480 & 0.606 & 0.585 & 0.025 & 0.089 & 0.304 & 1.167 \\ 
  Student-$t$ & 0.991 & 0.959 & 1.112 & 1.054 & 1.224 & 2.438 & 1.326 & 0.995 \\ 
  Skew-$t$ & 0.932* & 1.009 & 1.112 & 1.010 & 1.077 & 1.659 & 1.031 & 0.926 \\ 
  OT & 0.900*** & 0.964* & 0.925*** & 1.008 & 1.013 & 0.981 & 0.919** & 0.948 \\ 
  MT & 0.901*** & 0.979* & 0.932*** & 1.014 & 1.010 & 0.995 & 0.943** & 0.969 \\ 
  OST & 0.899*** & 0.960* & 0.910*** & 0.993 & 1.005 & 0.969 & 0.903** & 0.931** \\ 
  MST & 0.908*** & 0.976* & 0.918*** & 0.996 & 1.001 & 0.982 & 0.929** & 0.954* \\  \hline
\rowcolor{light-gray}   Gaussian-SV & 0.416*** & 0.437*** & 0.568** & 0.573 & 0.026 & 0.088 & 0.288** & 1.121 \\ 
  Student-$t$-SV & 1.018 & 1.048 & 1.002 & 1.032 & 0.984 & 0.996 & 0.990 & 1.009 \\ 
  Skew-$t$-SV & 1.030 & 1.033 & 0.977 & 1.030 & 0.979 & 0.997 & 0.987 & 0.992 \\ 
  OT-SV & 1.011 & 1.032 & 0.973 & 1.035 & 0.973* & 0.989 & 0.979 & 0.991 \\ 
  MT-SV & 1.014 & 1.037 & 0.979 & 1.037 & 0.971** & 0.991 & 0.989 & 0.995 \\ 
  OST-SV & 1.018 & 1.032 & 0.968* & 1.020 & 0.976* & 0.989 & 0.976 & 0.983 \\ 
  MST-SV & 1.021 & 1.041 & 0.969* & 1.021 & 0.970** & 0.991 & 0.979 & 0.983 \\ \hline
& \multicolumn{4}{c}{(b) Inflation } & \multicolumn{4}{c}{(d) VIX} \\      \hline
\rowcolor{light-gray}   Gaussian & 0.077 & 0.124 & 0.127 & 0.103 & 0.041 & 0.093 & 0.131 & 0.152 \\ 
  Student-$t$ & 0.964 & 2.705 & 1.405 & 2.371 & 1.288 & 1.217 & 1.141 & 1.336 \\ 
  Skew-$t$ & 0.906*** & 2.177 & 1.361 & 2.172 & 1.149 & 1.203 & 1.035 & 1.288 \\ 
  OT & 0.941** & 1.019 & 0.993 & 1.133 & 1.000 & 1.054 & 0.966 & 1.011 \\ 
  MT & 0.941** & 1.008 & 0.995 & 1.126 & 1.008 & 1.044 & 0.961 & 1.001 \\ 
  OST & 0.939** & 1.012 & 0.997 & 1.122 & 1.017 & 1.040 & 0.949 & 1.001 \\ 
  MST & 0.938** & 1.005 & 0.994 & 1.123 & 1.018 & 1.037 & 0.951 & 1.000 \\ \hline
\rowcolor{light-gray}   Gaussian-SV & 0.074 & 0.123 & 0.119** & 0.109 & 0.040 & 0.094 & 0.122 & 0.151 \\ 
  Student-$t$-SV & 0.998 & 0.987 & 1.002 & 0.939* & 1.016 & 1.023 & 1.029 & 1.029 \\ 
  Skew-$t$-SV & 0.992 & 0.988 & 0.998 & 0.970 & 1.027 & 1.021 & 1.020 & 1.018 \\ 
  OT-SV & 0.991 & 1.002 & 1.017 & 0.973 & 1.017 & 1.028 & 1.035 & 1.027 \\ 
  MT-SV & 0.991 & 0.999 & 1.015 & 0.976 & 1.018 & 1.025 & 1.029 & 1.023 \\ 
  OST-SV & 0.997 & 0.998 & 1.011 & 0.958 & 1.032 & 1.021 & 1.012 & 0.999 \\ 
  MST-SV & 0.995 & 0.996 & 1.014 & 0.956 & 1.030 & 1.021 & 1.011 & 1.001 \\ 
   \hline
\end{tabular}
\begin{tablenotes}
      \item {\small       
      Each panel reports the MSFE of the models relative to the Gaussian VAR model with (and without) stochastic volatility. The relative improvements over the Gaussian models are computed as the ratio of the MSFE of alternative specifications over the Gaussian models during 2007-2019. As the VAR models are nested, we perform a one-sided \cite{Diebold1995} test where the standard errors of the test statistics are computed with the Newey–West estimator \citep{Clark2011}. ***,**,* denote that the corresponding model significantly outperforms the Gaussian VAR at 1\%, 5\%, 10\% level. 
       The entries less than 1 indicate that the given model is better.}
    \end{tablenotes}
  \end{threeparttable}  
}
\end{table}

Table \ref{logscoreRecursive} reports the relative improvements in LP over the Gaussian VAR models. Here entries greater than 0 indicate that the given model is better than the Gaussian non-SV/SV model. The findings for the density forecasts are consistent with the findings for the point forecasts for the non stochastic volatility group. In the VAR with stochastic volatility group, the density forecast of the heavy tail and skewness models is improved at all horizons for the industrial production growth, however it is only statistically significant for the 6 and 12 months ahead forecast. On the other hand, the short run forecast accuracy of VIX increases when including the heavy tail and skewness parameters. The density forecasts in OST and MST VAR models for unemployment rate and inflation rate remain similar to the Gaussian benchmark. 
In agreement with \cite{Clark2011} and \cite{Clark2015}, stochastic volatility in the Gaussian VAR model is a crucial characteristic to improve out-of-sample density forecast.

\begin{table}[!tbp] 
\centering
\caption{Improvement in LP over the Gaussian VAR models} \label{logscoreRecursive}
\resizebox{1\textwidth}{!}{%
\begin{threeparttable}
\begin{tabular}{*{1}{p{2.5cm}} *{8}{p{1.70cm}}}
  \hline
 & 1M & 3M & 6M & 12M & 1M & 3M & 6M & 12M \\   \hline
&  \multicolumn{4}{c}{(a) Industrial Production} & \multicolumn{4}{c}{(c) Unemployment rate} \\  \hline
\rowcolor{light-gray} Gaussian & -1.044 & -1.122 & -1.236 & -1.266 & -0.075 & -0.600 & -1.072 & -1.712 \\ 
  Student-$t$ & 0.066*** & 0.056*** & 0.100*** & 0.073*** & 0.007** & -0.158 & -0.263 & -0.000 \\ 
  Skew-$t$ & 0.060** & 0.044** & 0.089*** & 0.078*** & 0.004 & -0.096 & -0.138 & -0.010 \\ 
  OT & 0.063** & 0.048* & 0.046* & 0.040** & -0.003 & -0.043 & -0.076 & -0.012 \\ 
  MT & 0.060** & 0.037* & 0.039* & 0.038** & 0.000 & -0.048 & -0.087 & -0.021 \\ 
  OST & 0.062** & 0.052** & 0.057** & 0.058*** & -0.002 & -0.037 & -0.067 & 0.007 \\ 
  MST & 0.057** & 0.042* & 0.051** & 0.054*** & 0.001* & -0.041 & -0.072 & 0.005 \\ \hline
\rowcolor{light-gray}  Gaussian-SV & -0.895*** & -0.936*** & -1.098* & -1.135 & 0.420*** & -0.139*** & -0.669*** & -1.494 \\ 
  Student-$t$-SV & 0.001 & -0.001 & 0.036** & 0.028 & -0.008 & -0.014 & -0.049 & -0.099 \\ 
  Skew-$t$-SV & -0.007 & 0.004 & 0.041*** & 0.043* & -0.006 & -0.019 & -0.076 & -0.047 \\ 
  OT-SV & 0.005 & 0.020 & 0.060** & 0.047** & -0.002 & -0.011 & -0.053 & -0.024 \\ 
  MT-SV & 0.008 & 0.016 & 0.062* & 0.050** & -0.002 & -0.014 & -0.050 & -0.074 \\ 
  OST-SV & 0.008 & 0.019 & 0.050** & 0.037* & -0.001 & 0.004 & -0.034 & -0.071 \\ 
  MST-SV & 0.002 & 0.013 & 0.053** & 0.052** & -0.002 & -0.020 & -0.054 & -0.086 \\ \hline
& \multicolumn{4}{c}{(b) Inflation } & \multicolumn{4}{c}{(d) VIX} \\       \hline
\rowcolor{light-gray} Gaussian & -0.372 & -0.591 & -0.647 & -0.665 & -0.139 & -0.523 & -0.697 & -0.860 \\ 
  Student-$t$ & 0.051*** & -0.024 & 0.041*** & -0.122 & 0.036*** & 0.016 & 0.071** & 0.101** \\ 
  Skew-$t$ & 0.048*** & 0.000 & 0.041*** & -0.089 & 0.045*** & 0.027** & 0.089*** & 0.110** \\ 
  OT & 0.046*** & 0.012 & -0.033 & -0.078 & 0.025*** & -0.028 & -0.034 & -0.005 \\ 
  MT & 0.046*** & 0.012 & -0.033 & -0.077 & 0.025*** & -0.025 & -0.032 & 0.001 \\ 
  OST & 0.046*** & 0.015 & -0.028 & -0.071 & 0.053*** & 0.044*** & 0.054*** & 0.090*** \\ 
  MST & 0.047*** & 0.016 & -0.027 & -0.070 & 0.054*** & 0.045*** & 0.056*** & 0.095*** \\ \hline
\rowcolor{light-gray}   Gaussian-SV & 0.056*** & -0.214*** & -0.242*** & -0.215*** & 0.184*** & -0.252*** & -0.423*** & -0.513*** \\ 
  Student-$t$-SV & -0.005 & 0.014 & 0.004 & 0.014 & 0.030* & -0.008 & -0.018 & -0.005 \\ 
  Skew-$t$-SV & -0.008 & -0.004 & 0.003 & -0.011 & 0.073*** & 0.020 & 0.012 & 0.005 \\ 
  OT-SV & -0.003 & -0.012 & 0.000 & -0.002 & 0.037** & -0.008 & -0.008 & -0.007 \\ 
  MT-SV & -0.004 & -0.013 & -0.002 & -0.004 & 0.037** & -0.007 & -0.004 & -0.006 \\ 
  OST-SV & -0.001 & -0.007 & 0.001 & 0.007 & 0.080*** & 0.046* & 0.037 & 0.010 \\ 
  MST-SV & -0.004 & -0.007 & 0.001 & 0.007 & 0.079*** & 0.038 & 0.037 & 0.010 \\ 
   \hline
\end{tabular}
\begin{tablenotes}
      \item {\small  
      Each panel reports the LP of the models relative to the Gaussian VAR model with (and without) stochastic volatility. The relative improvements over the Gaussian models are computed as the difference between the LP of alternative specifications and the Gaussian models during 2007-2019. As the VAR models are nested, we perform a one-sided \cite{Diebold1995} test where the standard errors of the test statistics are computed with the Newey–West estimator \citep{Clark2011}. ***,**,* denote that the corresponding model significantly outperforms the Gaussian VAR at 1\%, 5\%, 10\% level. The entries greater than 0 indicate that the given model is better.}
    \end{tablenotes}
  \end{threeparttable}  
}
\end{table}

Table \ref{CRPS} reports the relative improvements in CRPS over the Gaussian VAR models where entries greater than 0 indicate that the given model is better. 
We confirm the previous conclusion by comparing the CRPS among models. However, the effect of heavy tails and skewness is smaller as the CRPS is less sensitive to outliers \citep{Clark2015}. Skewness and heavy tailed VAR models with stochastic volatility improve significantly on the Gaussian model with SV in the medium term forecast of industrial production and the short term for the unemployment rate. The assumption of stochastic volatility is still essential for the density forecasts.

\begin{table}[!htbp] 
\centering
\caption{Improvement in CRPS over the Gaussian VAR models} \label{CRPS}
\resizebox{1\textwidth}{!}{%
\begin{threeparttable}
\begin{tabular}{*{1}{p{2.5cm}} *{8}{p{1.7cm}}}
  \hline
 & 1M & 3M & 6M & 12M & 1M & 3M & 6M & 12M \\   \hline
&  \multicolumn{4}{c}{(a) Industrial Production} & \multicolumn{4}{c}{(c) Unemployment rate} \\      \hline
\rowcolor{light-gray}   Gaussian & -0.378 & -0.376 & -0.431 & -0.442 & -0.116 & -0.198 & -0.346 & -0.666 \\ 
  Student-$t$ & 0.007 & -0.008 & -0.028 & 0.007 & -0.006 & -0.138 & -0.092 & 0.010 \\ 
  Skew-$t$ & 0.029*** & -0.011 & -0.021 & 0.008 & -0.001 & -0.082 & -0.036 & 0.025** \\ 
  OT & 0.030*** & 0.006* & 0.020*** & 0.020*** & -0.002 & -0.011 & -0.006 & 0.018** \\ 
  MT & 0.029*** & 0.003 & 0.018*** & 0.018*** & -0.002 & -0.011 & -0.010 & 0.010 \\ 
  OST & 0.030*** & 0.009** & 0.026*** & 0.025*** & -0.002 & -0.009 & -0.003 & 0.027*** \\ 
  MST & 0.029*** & 0.005** & 0.022*** & 0.025*** & -0.002 & -0.011 & -0.005 & 0.022*** \\  \hline
\rowcolor{light-gray}  Gaussian-SV & -0.331*** & -0.337*** & -0.390*** & -0.397** & -0.092*** & -0.159*** & -0.270*** & -0.536** \\ 
  Student-$t$-SV & -0.002 & -0.011 & -0.000 & 0.004 & 0.001 & -0.001 & 0.001 & -0.003 \\ 
  Skew-$t$-SV & -0.003 & -0.006 & 0.009 & 0.003 & 0.001 & -0.000 & -0.001 & -0.000 \\ 
  OT-SV & -0.001 & -0.006 & 0.011* & 0.003 & 0.002** & 0.000 & 0.000 & 0.001 \\ 
  MT-SV & -0.002 & -0.007 & 0.010* & 0.003 & 0.002** & 0.001 & -0.000 & 0.001 \\ 
  OST-SV & -0.002 & -0.005 & 0.014** & 0.005 & 0.002** & 0.001 & 0.001 & 0.001 \\ 
  MST-SV & -0.002 & -0.007 & 0.013** & 0.008 & 0.002** & 0.001 & 0.000 & -0.001 \\  \hline
& \multicolumn{4}{c}{(b) Inflation } & \multicolumn{4}{c}{(d) VIX} \\   \hline           
\rowcolor{light-gray}   Gaussian & -0.176 & -0.208 & -0.219 & -0.206 & -0.132 & -0.195 & -0.239 & -0.270 \\ 
  Student-$t$ & 0.016*** & -0.163 & -0.035 & -0.130 & -0.012 & -0.012 & -0.008 & -0.021 \\ 
  Skew-$t$ & 0.018*** & -0.116 & -0.031 & -0.108 & -0.007 & -0.013 & 0.006 & -0.013 \\ 
  OT & 0.011*** & -0.003 & -0.005 & -0.026 & 0.002** & -0.006 & 0.002 & 0.001 \\ 
  MT & 0.011*** & -0.003 & -0.004 & -0.025 & 0.003*** & -0.005 & 0.002 & 0.002 \\ 
  OST & 0.012*** & -0.002 & -0.003 & -0.024 & 0.004*** & 0.003* & 0.016*** & 0.017*** \\ 
  MST & 0.012*** & -0.002 & -0.004 & -0.024 & 0.005*** & 0.004** & 0.015*** & 0.018*** \\   \hline
\rowcolor{light-gray}   Gaussian-SV & -0.136*** & -0.172*** & -0.174*** & -0.168*** & -0.108*** & -0.170*** & -0.200*** & -0.223*** \\ 
  Student-$t$-SV & -0.002 & -0.000 & 0.002 & 0.005 & 0.001 & 0.002 & 0.000 & -0.001 \\ 
  Skew-$t$-SV & 0.000 & -0.000 & 0.003** & 0.001 & 0.001 & 0.001 & 0.002 & 0.000 \\ 
  OT-SV & 0.001 & -0.001 & 0.001 & 0.000 & 0.001 & 0.001 & 0.002 & 0.000 \\ 
  MT-SV & 0.001 & -0.001 & 0.001 & 0.000 & 0.001 & 0.001 & 0.002 & 0.001 \\ 
  OST-SV & 0.001 & -0.000 & 0.001 & 0.002 & 0.000 & 0.001 & 0.003 & 0.003 \\ 
  MST-SV & 0.001 & -0.000 & 0.001 & 0.002 & 0.000 & 0.001 & 0.003 & 0.002 \\ 
   \hline
\end{tabular}
\begin{tablenotes}
      \item {\small       
      Each panel reports the CRPS of the models relative to the Gaussian VAR model with (and without) stochastic volatility. The relative improvements over the Gaussian models are computed as the difference between the CRPS of alternative specifications and the Gaussian models during 2007-2019. As the VAR models are nested, we perform a one-sided \cite{Diebold1995} test where the standard errors of the test statistics are computed with the Newey–West estimator \citep{Clark2011}. ***,**,* denote that the corresponding model significantly outperforms the Gaussian VAR at 1\%, 5\%, 10\% level. The entries greater than 0 indicate that the given model is better.}
    \end{tablenotes}
  \end{threeparttable}  
}
\end{table}

\begin{figure}[!htbp] 
    \includegraphics[width=1\textwidth]{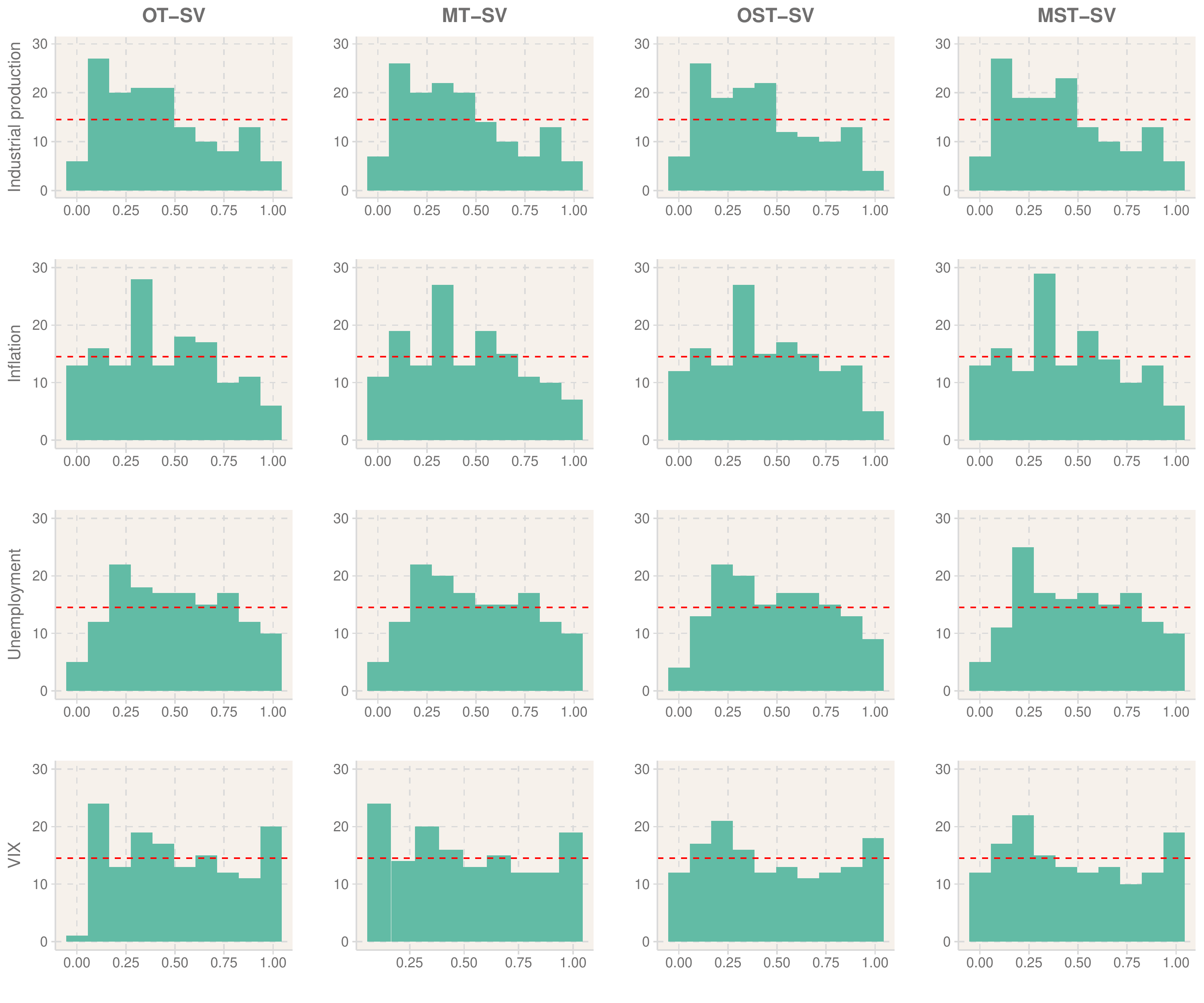}
    \caption{PIT histograms at the three month ahead forecast horizon and models with fat tails and skewness.}
    \label{fig:PIT}
    \caption*{{\scriptsize } }
\end{figure}

Next, we concentrate on the effect of skewness parameters in VAR models with stochastic volatility. Figure \ref{fig:PIT} shows the probability integral transforms (PITs) of the three month ahead forecast horizon from OT, MT, OST, MST VAR models. 
Without significant evidence of skewness (see Figure \ref{fig:NuSVvsnonSV}), the PIT plots for industrial production, inflation and unemployment are very similar. For the VIX the OT-SV model clearly overestimates the left tail of VIX with only a few observations classified as extreme according to the density forecast. The MT VAR model on the other hand has a slight tendency to overestimate the left tail of the VIX. The OST and MST specifications perform better here and we see that allowing for skewness can make a difference. The result is similar for the PIT plots at the other horizons.

\newpage

\begin{figure}[!htbp]  
\begin{center}
    \includegraphics[width=0.9\textwidth]{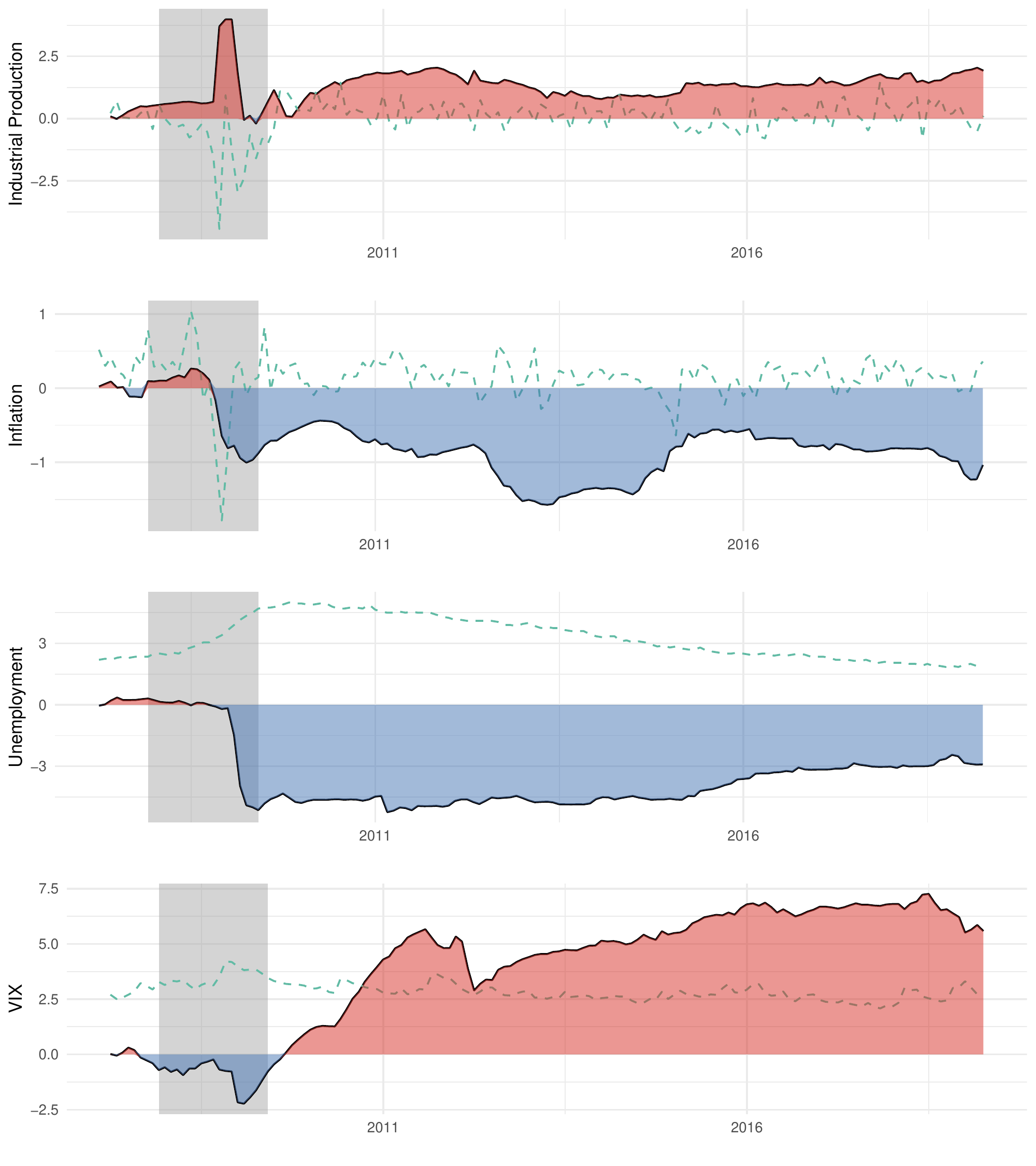}
    \caption{Cumulative log Bayes factors of the predictive density for 3 month ahead forecasts between the Gaussian-SV and MST-SV models.}
    \label{fig:cumLogBF}	
    \caption*{{\scriptsize Positive values (red) means MST-SV predicts better and negative values (blue) means that Gaussian model does better. The dashed lines illustrate the scale values of the original variables. See \cite{Geweke2010} for details.} }
\end{center}
\end{figure}

Figure \ref{fig:cumLogBF} shows the cumulative log Bayes factors of the predictive density for 3 month forecast horizon between the Gaussian-SV and MST-SV models, see the computational details in \cite{Geweke2010}. 
Positive values (red) means that MST-SV predicts better than the Gaussian-SV.
A common feature across the variables is that the MST-SV performs better than or roughly on par with the Gaussian-SV up to the middle of the recession and performs worse close to the end. For industrial production the MST-SV performs better overall and regains it advantage after the recession. The Gaussian-SV performs better overall for inflation and unemployment but it is noteworthy that, for unemployment, the MST-SV consistently outperforms the Gaussian-SV by a small margin over the expansion. For the VIX the MST-SV also improves its performance during the expansion and does significantly better overall. This pattern suggests that allowing for heavy tails and skewness is not just about accounting for large deviations but also about modelling the more central parts of the distribution well and that the latter can be equally important. Recalling that \cite{Chiu2017} interpreted the mixing variables as accounting for high frequency shocks we can also see this as a factor explaining the overall better performance of the MST-SV during the expansion.

%
%

\section{Conclusion}
 Skewness and heavy tails are empirically relevant features in many application areas -- not only the macroeconomic and financial application we consider in this paper. While these features to some extent can be accommodated or masked by time-varying heteroskedasticity modelled as GARCH-type or stochastic volatility processes there is a need for models that explicitly account for skewness and heavy tails in the data.
We contribute to this by proposing flexible skew and heavy tailed distributions with the symmetric normal distribution as a special case. Specifically, we introduce a general class of Generalized Hyperbolic Skew Student's $t$ distributions with stochastic volatility for VAR models. The stochastic representation of the GHSkew-$t$ can be written in term of a variance-mean mixture which leads to a straightforward implementation of a Gibbs sampler for posterior inference. We also take advantage of the cross entropy methods by \cite{Chan2018} to calculate the model marginal likelihood and compare the in-sample fit among different specifications. In an application to US data we find support for VAR models with skewness and heavy tails. The VAR models with skewness and heavy tails gives better point forecasts and density forecasts compared to Gaussian VAR models for many, but not all, variables we model. 
We recommend that skewness should be taken into account for improving  forecasting performance during recessions and crises.

\newpage

\section*{Acknowledgment}

We thank Pär Österholm for helpful comments. The authors acknowledge financial support from the project ``Models for macro and financial economics after the financial crisis" (Dnr: P18-0201, BV18-0018) funded by the Jan Wallander and Tom Hedelius Foundation. Stepan Mazur also acknowledges financial support from the internal research grants of Örebro University. The computations were enabled by resources provided by the Swedish National Infrastructure for Computing (SNIC) at HPC2N partially funded by the Swedish Research Council through grant agreement no. 2018-05973.

\appendix

\begin{center}
{\Large \textbf{Appendix}}
\end{center}

\section{The integrated likelihood} \label{intoverHW}
The integrated likelihood $p(y_{1:T}|\theta_1)$ require a high dimensional integral over the latent states $\boldsymbol \theta_2 = \{ \mathbf \xi_{1:K,1:T}^{'}, \mathbf h_{1:K,1:T}^{'} \}^{'}$,
\begin{equation*}
\begin{aligned}
p(y_{1:T}|\theta_1) & = \int \int p(y_{1:T}|\theta_1,  \mathbf \xi_{1:K,1:T}, h_{1:K,1:T}) p(\mathbf \xi_{1:K,1:T}, h_{1:K,1:T} |\theta_1 ) d \mathbf \xi_{1:K,1:T} d h_{1:K,1:T}.
\end{aligned}
\end{equation*}
The integral can be solved by an importance sampling step over $h_{1:K,1:T}$ or over $\mathbf \xi_{1:K,1:T}$,
\begin{equation*}
\begin{aligned}
(A1) \;\;\;\; p(y_{1:T}|\theta_1) & = \int p(y_{1:T}|\theta_1, h_{1:K,1:T}) p(h_{1:K,1:T} | \theta_1 ) d h_{1:K,1:T} \\
& \approx \sum_{l = 1}^{L} \frac{1}{L}  \frac{p(y_{1:T}|\theta_1, h_{1:K,1:T}^{(l)} ) p(h_{1:K,1:T}^{(l)} | \theta_1 ) }{f(h_{1:K,1:T}^{(l)} | \lambda_H) } \\
(A2) \;\;\;\; p(y_{1:T}|\theta_1) & = \int p(y_{1:T}|\theta_1,  \mathbf \xi_{1:K,1:T}) p(\mathbf \xi_{1:K,1:T} | \theta_1 ) d \mathbf \xi_{1:K,1:T} \\
& \approx \sum_{m = 1}^{M} \frac{1}{M}  \frac{p(y_{1:T}|\theta_1, \mathbf \xi_{1:K,1:T}^{(m)} ) p(\mathbf \xi_{1:K,1:T}^{(m)} | \theta_1 )}{f(\mathbf \xi_{1:K,1:T}^{(m)} | \lambda_W)}
\end{aligned}
\end{equation*}
(A1) proposes an importance sampling distribution $f(h_{1:K,1:T} | \lambda_H)$ and simulate $h_{1:K,1:T}^{(l)} \sim f(h_{1:K,1:T} | \lambda_H) $ for $l = 1, \ldots, L$. Then, $p(y_{1:T}|\theta_1, h_{1:K,1:T}^{(l)})$ is the conditional likelihood which can be derived in a closed form as multivariate Gaussian, multivariate Student-$t$, multivariate hyperbolic skew Student-$t$, orthogonal Student-$t$, orthogonal hyperbolic skew Student-$t$ for the Gaussian, Student-$t$, Skew-$t$, OT, OST VAR models respectively. The MT and MST VAR models does not have a closed form expression for $p(y_{1:T}|\theta_1, h_{1:K,1:T}^{(l)})$ for these $p(y_{1:T}|\theta_1, h_{1:K,1:T}^{(l)})$ is estimated using importance sampling with the Metropolis-Hasting proposal from step 7 of the Gibbs sampler as the importance function. Following \cite{Chan2018} we take the importance function  $f(h_{1:K,1:T} | \lambda_H)$ to be a multivariate normal distribution. The importance sampling mean and precision matrix $\lambda_H = \{ \hat{h}_{1:K,1:T}, \hat{\Sigma}^{-1}_{H} \}$ can be chosen as
\begin{equation*}
\begin{aligned}
\hat{h}_{1:K,1:T} & = \argmax_{h_{1:K,1:T}} \log p(h_{1:K,1:T} |y_{1:T}, \theta_1), \\
\hat{\Sigma}^{-1}_{H} & = - \frac{\partial^2 \log p(h_{1:K,1:T} |y_{1:T}, \theta_1) }{ \partial h_{1:K,1:T} ^2 } \Bigg|_{h_{1:K,1:T} = \hat{h}_{1:K,1:T}}.
\end{aligned}
\end{equation*}
We have that $p(h_{1:K,1:T} |y_{1:T}, \theta_1) \propto p(y_{1:T} | h_{1:K,1:T},  \theta_1) p( h_{1:K,1:T} | \theta_1)$. As the derivative of $p(y_{1:T} | h_{1:K,1:T},  \theta_1)$ is computationally expensive, we approximate it by fixing $\mathbf \xi_{1:K,1:T}$ at the posterior mean. 

(A2) proposes an importance sampling distribution $f(\mathbf \xi_{1:K,1:T} | \lambda_W)$ and simulate $\mathbf \xi_{1:K,1:T}^{(m)} \sim f(\mathbf \xi_{1:K,1:T} | \lambda_W) $ for $m = 1, \ldots, M$. Then, $p(y_{1:T}|\theta_1, \mathbf \xi_{1:K,1:T}^{(m)})$ is the conditional likelihood which can be derived as a Gaussian multivariate distribution.
\cite{Chan2018} show a good example for calculating the conditional likelihood $p(y_{1:T}|\theta_1, \mathbf \xi_{1:K,1:T}^{(m)})$ in a Gaussian case. 
It is, however, difficult to come up with a good importance function $f(\mathbf \xi_{1:K,1:T}| \lambda_W)$. 
One possibility is the Metropolis Hasting proposal distribution of $\mathbf \xi_{1:K,1:T}$ in the Gibbs sampling scheme (Step 7). We  thus we sample $\mathbf \xi_{1:K,1:T}^{(1)}, \ldots, \mathbf \xi_{1:K,1:T}^{(M)} \sim p_{MH}(\mathbf \xi_{1:K,1:T} | y_{1:T},\theta_1, \bar{H}_{1:T})$, where $\bar{H}_{1:T}$ is the posterior mean of $h_{1:K,1:T}$.

We note that with the same number of importance samples L and M, the integrated likelihood estimated by (A1) gives a smaller variance in comparison to that by (A2). It is not only due to the closed form expression of $p(y_{1:T}|\theta_1, h_{1:K,1:T})$ but also due to $p(y_{1:T}|\theta_1, h_{1:K,1:T}) = \displaystyle\prod_{t=1}^T p(y_{t}|\theta_1,  h_{1:K,t}, y_{1:t-1})$. So the integral can be separated for each $W_t$ and hence be more accurate. Table \ref{LML} utilizes (A1) for the integrated likelihood.

\newpage

\section{Posterior comparison} \label{PostApp}

\begin{figure}[!htbp] 
    \includegraphics[width=1\textwidth]{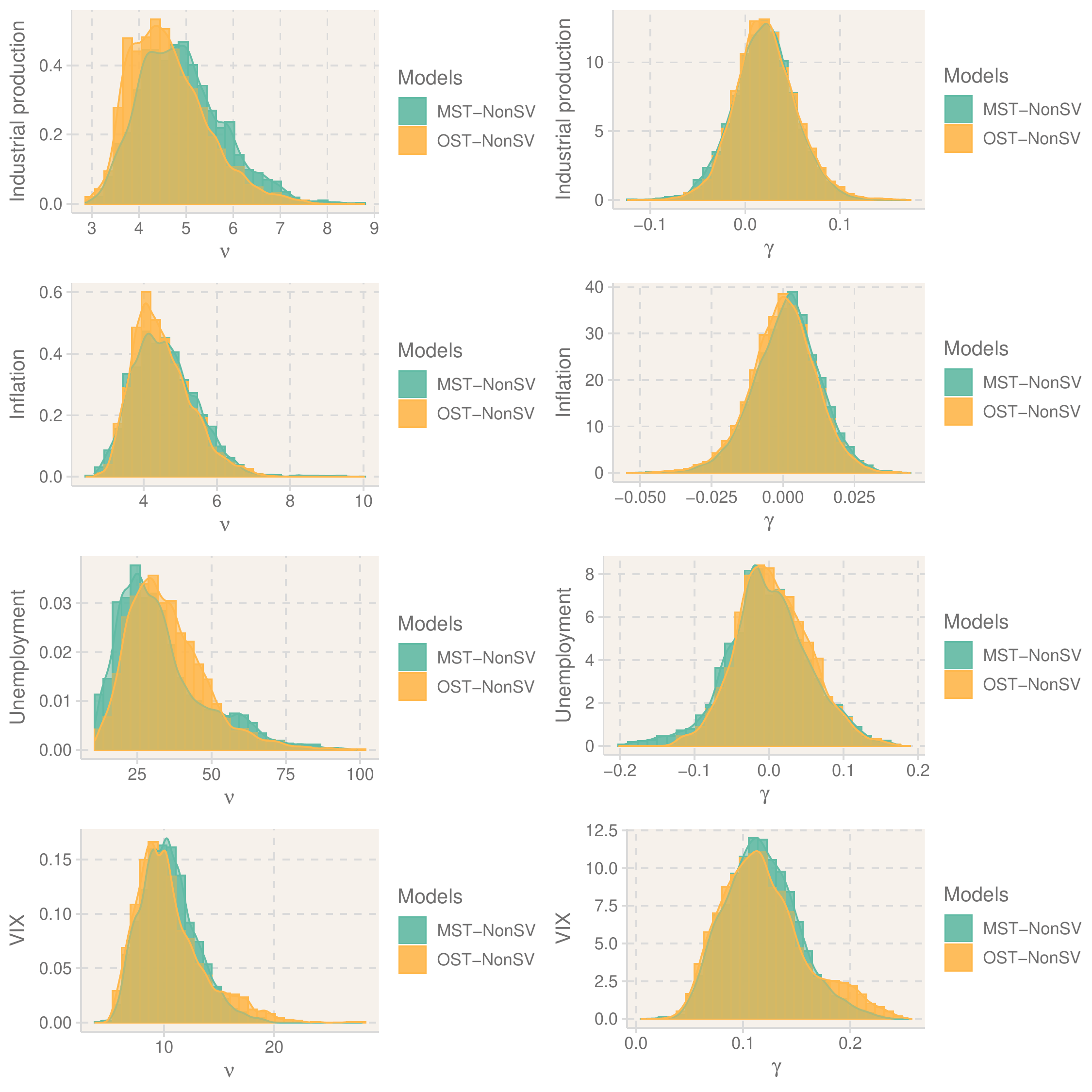}
    \caption{The plots show the posterior samples of the heavy tail and skewness parameters of the MST VAR model and OST VAR model without SV.}
    \label{fig:samplesbifcop2}
    \caption*{  }
\end{figure}
\newpage

\begin{figure}[!htbp] 
    \includegraphics[width=1\textwidth]{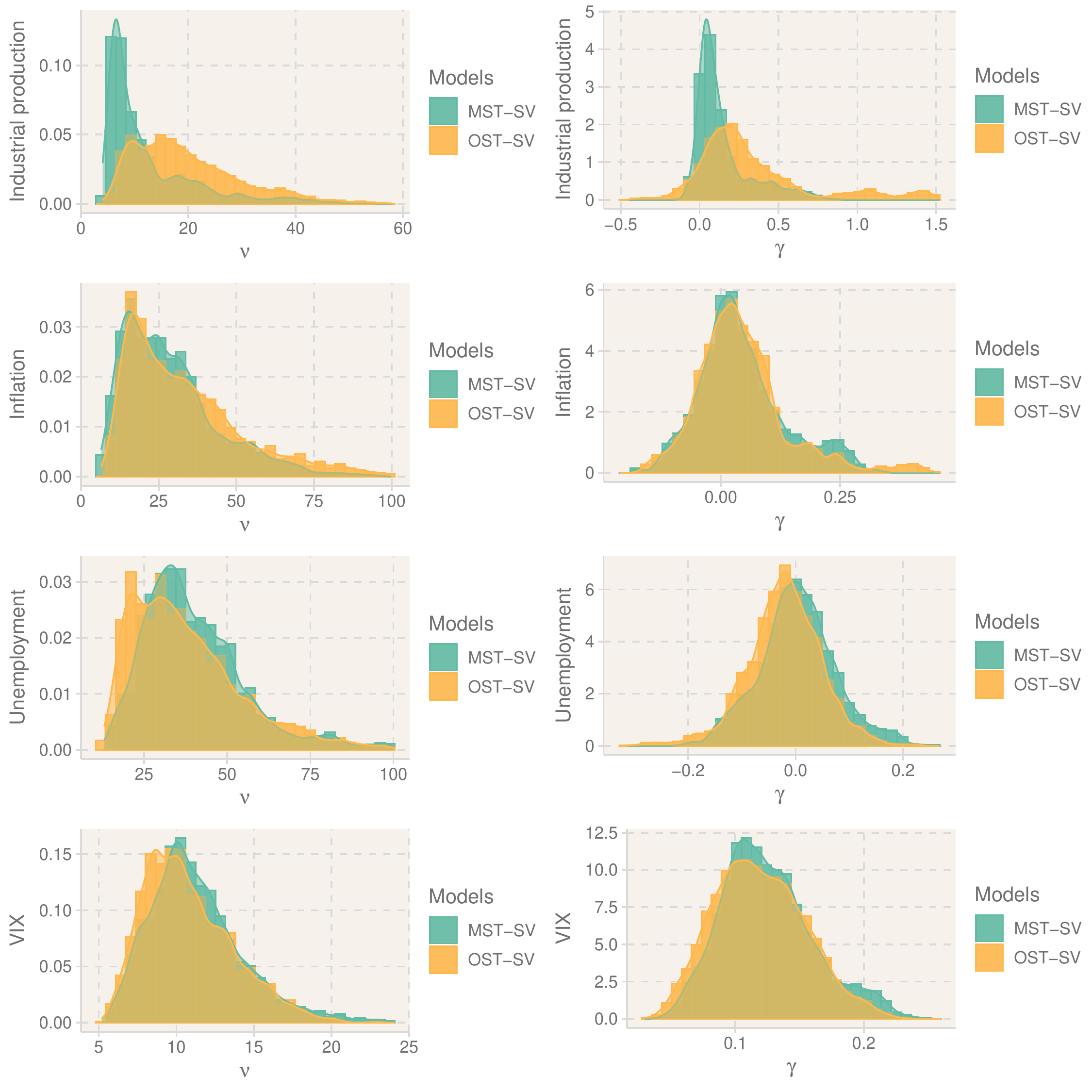}
    \caption{The plots show the posterior samples of the heavy tail and skewness parameters of the MST VAR model and OST VAR model with SV.}
    \label{fig:samplesbifcop3}
    \caption*{  }
\end{figure}

\newpage

\bibliographystyle{apalike}
\bibliography{WP5}

\begin{thebibliography}{}

\bibitem[Aas and Haff, 2006]{Aas2006}
Aas, K. and Haff, I.~H. (2006).
\newblock The generalized hyperbolic skew {Student's} t-distribution.
\newblock {\em Journal of Financial Econometrics}, 4(2):275--309.

\bibitem[Acemoglu et~al., 2017]{Acemoglu2017}
Acemoglu, D., Ozdaglar, A., and Tahbaz-Salehi, A. (2017).
\newblock Microeconomic origins of macroeconomic tail risks.
\newblock {\em American Economic Review}, 107(1):54--108.

\bibitem[Andersson and Karlsson, 2008]{Andersson2008}
Andersson, M.~K. and Karlsson, S. (2008).
\newblock Bayesian forecast combination for {VAR} models.
\newblock In Chib, S. and Griffiths, W., editors, {\em Bayesian econometrics},
  volume~23, pages 501--524. Emerald Group Publishing.

\bibitem[Carriero et~al., 2021]{Carriero2021}
Carriero, A., Clark, T.~E., Marcellino, M., and Mertens, E. (2021).
\newblock Addressing {COVID-19 Outliers in BVARs with Stochastic Volatility}.
\newblock Working Papers 21-02, Federal Reserve Bank of Cleveland.

\bibitem[Carriero et~al., 2020]{Carriero2020}
Carriero, A., Clark, T.~E., and Marcellino, M.~G. (2020).
\newblock Capturing macroeconomic tail risks with {Bayesian} vector
  autoregressions.
\newblock Working Papers 20-02R, Federal Reserve Bank of Cleveland.

\bibitem[Carter and Kohn, 1994]{Carter1994}
Carter, C.~K. and Kohn, R. (1994).
\newblock On gibbs sampling for state space models.
\newblock {\em Biometrika}, 81(3):541--553.

\bibitem[Chan and Eisenstat, 2018]{Chan2018}
Chan, J.~C. and Eisenstat, E. (2018).
\newblock Bayesian model comparison for time-varying parameter {VARs} with
  stochastic volatility.
\newblock {\em Journal of Applied Econometrics}, 33(4):509--532.

\bibitem[Chib and Ramamurthy, 2014]{Chib2014}
Chib, S. and Ramamurthy, S. (2014).
\newblock {DSGE Models with Student-t errors}.
\newblock {\em Econometric Reviews}, 33(1-4):152--171.

\bibitem[Chiu et~al., 2017]{Chiu2017}
Chiu, C.-W.~J., Mumtaz, H., and Pinter, G. (2017).
\newblock Forecasting with {VAR} models: {F}at tails and stochastic volatility.
\newblock {\em International Journal of Forecasting}, 33(4):1124--1143.

\bibitem[Christiano, 2007]{Christiano2007}
Christiano, L.~J. (2007).
\newblock Comment {[On the Fit of New Keynesian Models]}.
\newblock {\em Journal of Business {\&} Economic Statistics}, 25(2):143--151.

\bibitem[Clark, 2011]{Clark2011}
Clark, T.~E. (2011).
\newblock Real-time density forecasts from {Bayesian} vector autoregressions
  with stochastic volatility.
\newblock {\em Journal of Business \& Economic Statistics}, 29(3):327--341.

\bibitem[Clark and Ravazzolo, 2015]{Clark2015}
Clark, T.~E. and Ravazzolo, F. (2015).
\newblock Macroeconomic forecasting performance under alternative
  specifications of time-varying volatility.
\newblock {\em Journal of Applied Econometrics}, 30(4):551--575.

\bibitem[Cogley and Sargent, 2005]{Cogley2005}
Cogley, T. and Sargent, T.~J. (2005).
\newblock Drifts and volatilities: {M}onetary policies and outcomes in the post
  {WWII US}.
\newblock {\em Review of Economic Dynamics}, 8(2):262--302.

\bibitem[Creal and Tsay, 2015]{Creal2015}
Creal, D.~D. and Tsay, R.~S. (2015).
\newblock High dimensional dynamic stochastic copula models.
\newblock {\em Journal of Econometrics}, 189(2):335--345.

\bibitem[Cross and Poon, 2016]{Cross2016}
Cross, J. and Poon, A. (2016).
\newblock Forecasting structural change and fat-tailed events in {Australian}
  macroeconomic variables.
\newblock {\em Economic Modelling}, 58:34--51.

\bibitem[C{\'u}rdia et~al., 2014]{Curdia2014}
C{\'u}rdia, V., Del~Negro, M., and Greenwald, D.~L. (2014).
\newblock Rare shocks, great recessions.
\newblock {\em Journal of Applied Econometrics}, 29(7):1031--1052.

\bibitem[Del~Negro and Primiceri, 2015]{Del2015}
Del~Negro, M. and Primiceri, G.~E. (2015).
\newblock Time varying structural vector autoregressions and monetary policy:
  {A} corrigendum.
\newblock {\em The Review of Economic Studies}, 82(4):1342--1345.

\bibitem[Delle~Monache et~al., 2020]{Delle2020}
Delle~Monache, D., De~Polis, A., and Petrella, I. (2020).
\newblock Modeling and forecasting macroeconomic downside risk.
\newblock Research Papers~34, Economic Modelling and Forecasting Group,
  University of Warwick.

\bibitem[Diebold and Mariano, 1995]{Diebold1995}
Diebold, F.~X. and Mariano, R.~S. (1995).
\newblock Comparing predictive accuracy.
\newblock {\em Journal of Business \& economic statistics}, 13(3):134--144.

\bibitem[Fagiolo et~al., 2008]{Fagiolo2008}
Fagiolo, G., Napoletano, M., and Roventini, A. (2008).
\newblock Are output growth-rate distributions fat-tailed? some evidence from
  {OECD} countries.
\newblock {\em Journal of Applied Econometrics}, 23(5):639--669.

\bibitem[Ferreira and Steel, 2007]{Ferreira2007}
Ferreira, J.~T. and Steel, M.~F. (2007).
\newblock A new class of skewed multivariate distributions with applications to
  regression analysis.
\newblock {\em Statistica Sinica}, pages 505--529.

\bibitem[Geweke and Amisano, 2010]{Geweke2010}
Geweke, J. and Amisano, G. (2010).
\newblock Comparing and evaluating {Bayesian} predictive distributions of asset
  returns.
\newblock {\em International Journal of Forecasting}, 26(2):216--230.

\bibitem[Gneiting and Raftery, 2007]{Gneiting2007}
Gneiting, T. and Raftery, A.~E. (2007).
\newblock Strictly proper scoring rules, prediction, and estimation.
\newblock {\em Journal of the American statistical Association},
  102(477):359--378.

\bibitem[Karlsson, 2013]{Karlsson2013}
Karlsson, S. (2013).
\newblock Forecasting with {Bayesian} vector autoregression.
\newblock In Elliott, G. and Timmermann, A., editors, {\em Handbook of economic
  forecasting}, volume~2, pages 791--897. Elsevier.

\bibitem[Karlsson and Mazur, 2020]{Karlsson2020}
Karlsson, S. and Mazur, S. (2020).
\newblock Flexible fat-tailed vector autoregression.
\newblock Working Papers 2020:5, Örebro University, School of Business.

\bibitem[Kastner and Fr{\"u}hwirth-Schnatter, 2014]{Kastner2014}
Kastner, G. and Fr{\"u}hwirth-Schnatter, S. (2014).
\newblock Ancillarity-sufficiency interweaving strategy (asis) for boosting
  mcmc estimation of stochastic volatility models.
\newblock {\em Computational Statistics \& Data Analysis}, 76:408--423.

\bibitem[Kim et~al., 1998]{Kim1998}
Kim, S., Shephard, N., and Chib, S. (1998).
\newblock Stochastic volatility: {L}ikelihood inference and comparison with
  {ARCH} models.
\newblock {\em The Review of Economic Studies}, 65(3):361--393.

\bibitem[Koop and Korobilis, 2010]{Koop2010}
Koop, G. and Korobilis, D. (2010).
\newblock {\em Bayesian multivariate time series methods for empirical
  macroeconomics}.
\newblock Now Publishers Inc.

\bibitem[Liu, 2019]{Liu2019}
Liu, X. (2019).
\newblock On tail fatness of macroeconomic dynamics.
\newblock {\em Journal of Macroeconomics}, 62:103154.

\bibitem[McCracken and Ng, 2016]{McCracken2016}
McCracken, M.~W. and Ng, S. (2016).
\newblock {FRED-MD: A} monthly database for macroeconomic research.
\newblock {\em Journal of Business \& Economic Statistics}, 34(4):574--589.

\bibitem[McNeil et~al., 2015]{Mcneil2015}
McNeil, A.~J., Frey, R., and Embrechts, P. (2015).
\newblock {\em {Quantitative risk management}: {Concepts, Techniques and
  Tools-revised edition}}.
\newblock Princeton university press.

\bibitem[Nguyen et~al., 2019]{Nguyen2019}
Nguyen, H., Aus{\'\i}n, M.~C., and Galeano, P. (2019).
\newblock Parallel {Bayesian} inference for high-dimensional dynamic factor
  copulas.
\newblock {\em Journal of Financial Econometrics}, 17(1):118--151.

\bibitem[Ni and Sun, 2005]{Ni2005}
Ni, S. and Sun, D. (2005).
\newblock Bayesian estimates for vector autoregressive models.
\newblock {\em Journal of Business \& Economic Statistics}, 23(1):105--117.

\bibitem[Panagiotelis and Smith, 2008]{Panagiotelis2008}
Panagiotelis, A. and Smith, M. (2008).
\newblock Bayesian density forecasting of intraday electricity prices using
  multivariate skew-t distributions.
\newblock {\em International Journal of Forecasting}, 24(4):710--727.

\bibitem[Primiceri, 2005]{Primiceri2005}
Primiceri, G.~E. (2005).
\newblock Time varying structural vector autoregressions and monetary policy.
\newblock {\em The Review of Economic Studies}, 72(3):821--852.

\bibitem[Roberts and Rosenthal, 2009]{Roberts2009}
Roberts, G.~O. and Rosenthal, J.~S. (2009).
\newblock Examples of adaptive mcmc.
\newblock {\em {Journal of Computational and Graphical Statistics}},
  18(2):349--367.

\bibitem[Sahu et~al., 2003]{Sahu2003}
Sahu, S.~K., Dey, D.~K., and Branco, M.~D. (2003).
\newblock A new class of multivariate skew distributions with applications to
  {Bayesian} regression models.
\newblock {\em Canadian Journal of Statistics}, 31(2):129--150.

\bibitem[Sims, 1980]{Sims1980}
Sims, C.~A. (1980).
\newblock Macroeconomics and reality.
\newblock {\em Econometrica}, 48(1):1--48.

\bibitem[Uhlig, 1997]{Uhlig1997}
Uhlig, H. (1997).
\newblock Bayesian vector autoregressions with stochastic volatility.
\newblock {\em {Econometrica}}, pages 59--73.

\end{thebibliography}

\end{document}